\documentclass[fleqn,usenatbib]{mnras}

\usepackage{newtxtext,newtxmath}
\usepackage[T1]{fontenc}

\DeclareRobustCommand{\VAN}[3]{#2}
\let\VANthebibliography\thebibliography
\def\thebibliography{\DeclareRobustCommand{\VAN}[3]{##3}\VANthebibliography}

\usepackage{graphicx}	% Including figure files
\usepackage{amsmath}	% Advanced maths commands
\usepackage{anyfontsize}
\newcommand{\Mhalo}{$M_{\rm{200c}}$}    
\newcommand{\Msun}{$\mathrm{M_{\odot}}$}    
\title[LYRA ultra-faints]{LYRA ultra-faints: The emergence of faint dwarf galaxies in the presence of an early Lyman-Werner background}

\author[S. T. Brown et al.]{
Shaun T. Brown,$^{1,2}$\thanks{E-mail: shaun.t.brown@durham.ac.uk}
Azadeh Fattahi,$^{1,2}$
Thales A. Gutcke,$^{3}$
Sylvia Ploeckinger,$^{4}$
Joaquin Sureda,$^{1}$\newauthor
Sownak Bose,$^{1}$
Jessica E. Doppel,$^{1,5}$
R\"udiger Pakmor$^{6}$
and Adrian Jenkins$^{1}$
\\
$^{1}$Institute for Computational Cosmology, Durham University, South Road, Durham, DH1 3LE, UK\\
$^{2}$The Oskar Klein Centre, Department of Physics, Stockholm University, Albanova University Center, 106 91 Stockholm, Sweden\\
$^{3}$Institute for Astronomy, University of Hawaii, 2680 Woodlawn Drive, Honolulu, HI 96822, USA\\
$^{4}$Department of Astrophysics, University of Vienna, Türkenschanzstrasse 17, 1180 Vienna, Austria\\
$^{5}$Centre for Extragalactic Astronomy, Department of Physics, Durham University, Durham DH1 3LE, UK\\
$^{6}$Max-Planck-Institut f\"ur Astrophysik, Karl-Schwarzschild-Str. 1, D-85748, Garching, Germany
}

\date{Accepted XXX. Received YYY; in original form ZZZ}

\pubyear{\the\year{}}

\begin{document}
\label{firstpage}
\pagerange{\pageref{firstpage}--\pageref{lastpage}}
\maketitle

\begin{abstract}
We present a suite of zoom-in cosmological hydrodynamical simulations of dwarf galaxies using the LYRA galaxy formation model with an extremely high mass resolution of $4\, \mathrm{M_{\odot}}$, evolved to $z=0$. The suite contains 65 haloes selected from Local Group like environments, spanning $M_{\mathrm{200c}}=10^7$--$5\times10^9\, \mathrm{M_{\odot}}$. The sample includes small ultra-faints with $M_\ast\sim100\, \mathrm{M_{\odot}}$ through to classical dwarfs with $M_\ast \sim 5\times10^6  \mathrm{M_{\odot}}$, as well as haloes that remain dark to the present day. We explore two prescriptions for the high-redshift ($z>7$) Lyman-Werner background (LWB), differing in intensity and redshift evolution. Star formation begins early ($z\gtrsim8$) in progenitors with $M_{\mathrm{200c}}\sim10^5$–$10^6 \mathrm{M_{\odot}}$, where molecular hydrogen enables warm moderate-density gas to efficiently cool. The LWB strongly influences the $z=0$ halo occupation fraction, shifting the dark-to-luminous transition from $M_{\mathrm{200c}}\sim10^7 \mathrm{M_{\odot}}$ (weaker LWB) to $M_{\mathrm{200c}}\sim10^8 \mathrm{M_{\odot}}$ (stronger LWB). Galaxies with $M_\ast\gtrsim10^5  \mathrm{M_{\odot}}$ are mostly insensitive to the LWB choice, whereas lower-mass systems respond strongly, producing markedly different stellar mass-halo mass (SMHM) relations. The weaker LWB yields a very shallow SMHM slope with nearly constant scatter, while the stronger LWB introduces a pronounced break at $M_{\mathrm{200c}}\sim10^9 \mathrm{M_{\odot}}$, where haloes of similar mass host galaxies with $M_\ast\sim10^3$–$10^5 \mathrm{M_{\odot}}$ or remain dark. Both models produce a minimum stellar mass floor at $M_\ast\sim10^3 \mathrm{M_{\odot}}$, originating from galaxies that undergo a single burst of star formation at high redshift before self-quenching from their first supernovae.
\end{abstract}


%%%%%%%%%%%%%%%%%%%%%%%%%%%%%%%%%%%%%%%%%%%%%%%%%%

%%%%%%%%%%%%%%%%% BODY OF PAPER %%%%%%%%%%%%%%%%%%

\section{Introduction}

Observations of dwarf galaxies have made great progress in the past few decades. Starting from only a handful of known `classical' dwarf satellites in the early 2000s, the number has increased to approximately 65 confirmed Milky Way satellites today, with countless more identified in the field \cite[e.g.][]{Pace_24}. Modern photometric surveys have allowed for ever deeper observations that have unveiled a population of extremely low-mass ultra faint dwarf galaxies \cite[see][for a recent review]{dwarf_review}.

These small galaxies have proven to be powerful cosmological probes and the subject of an increasing number of studies. Being the most dark matter (DM) dominated known systems makes them natural laboratories to explore the nature of DM. Many of their properties, in particular the expected number of ultra-faints, are sensitive to the assumed dark matter model. Various alternatives to the standard cold DM paradigm have effects that manifest in the dwarf regime; examples include warm dark matter \citep[e.g.][]{Lovell_14, Stafford_20}, self-interacting DM \citep[e.g.][]{Kaplinghat, Victor_22, Gutcke2025} and `fuzzy', axionic, DM \cite[e.g.][]{Marsh_15, May_23}. Some of the tightest current constraints on alternative DM models come from dwarf galaxy satellite counts \cite[e.g.][]{Nadler_21_cons}.

Dwarf galaxies' shallow potentials also make them sensitive probes of feedback processes such as supernovae (SN) and stellar winds \citep[see][ for a review]{Collins_22}. This makes them ideal systems to study how feedback on small scales ($\sim \rm{pc}$) couples to galactic scale outflows ($\sim \rm{kpc}$) and help regulate star formation over cosmic time \citep[e.g.][]{Naab_17}. Furthermore, many dwarf galaxies exhibit extremely old stellar populations, often comparable to the age of the Universe \citep[e.g.][]{Tolstoy_09,Weisz_14}, meaning that these faint systems were likely formed prior to reionisation ($z \sim 7$) and represent local relics of the earliest phases of structure formation.

While observations have expanded our view of ultra-faints, modelling the formation and evolution of these small systems has proven a significant challenge. One of the most powerful tools to understand galaxy formation is the use of cosmological hydrodynamic simulations, which self-consistently evolve the distribution of matter from the early Universe through to the present day \citep[see][for a review]{Vogelsberger_review}. Many large volume cosmological simulations can now successfully reproduce realistic populations of galaxies that match a diverse range of observations (e.g. EAGLE \citep{EAGLE_1,EAGLE_2}, Illustris-TNG \citep{TNG1,TNG2}, Horizon-AGN \citep{Horizon-AGN}, Romulus \citep{Romulus} and Simba \citep{Simba}), and continue to improve their physical prescriptions; of particular note the recently published COLBRE simulations \citep{Colibre} which now resolve the cold phase of the ISM, the formation of molecular hydrogen and the production of dust. However, due to their computation cost these simulations are currently limited to minimum resolution elements of $\sim 10^5$--$10^6$\Msun, and lack the mass resolution needed to form the faintest dwarf galaxies.

An alternative approach to model galaxy formation is the use of semi-analytic models \citep[SAMs, e.g.][]{Galform, L-galaxies}. These have the advantage of having no minimum mass resolution, and are able to form galaxies over a large range of masses. This, combined with their modest computation cost, has made them particularly useful for modelling the Milky Way satellite population down to the ultra faint regime \citep[e.g.][]{Bose_20, Grumpy_model, Galacticus_2}. Various SAMs now successfully reproduce satellite number counts and many other of their intrinsic properties, such as their size and metallicities. However, as these types of models rely on the output of DM-only simulations, they have to make certain limiting assumptions about how small scale processes, such as feedback, couple to larger scales within the galaxies. This leads to these models being parametrised in flexible ways where the model parameters are often constrained to reproduce the observations. Nevertheless, these methods remain an invaluable tool for modelling and understanding the important physical processes giving rise to the observed population of dwarf galaxies.

Another promising approach to study the detailed formation of dwarf galaxies is to use dedicated zoom-in cosmological simulations of individual dwarf galaxies. Their low mass allows for them to be simulated at significantly higher resolutions than is currently  possible for cosmological volumes or Milky Way-mass zooms. A number of works in recent years have taken this approach to model isolated dwarf galaxies, achieving resolutions $\lesssim 100$\Msun. Recent examples include the EDGE project \citep{Edge_1, Edge_2}, subsets of the FIRE-2 simulations \citep{Fitts_17, Wheeler} and the GEAR simulations \citep{Gear}, and of particular relevance to this work the, LYRA simulations \citep{LYRA_1,LYRA_2} which use an extremely high resolution of $4$\Msun.

In this work we develop and extend the existing LYRA simulations by significantly expanding the size of the sample of haloes and covering a wider range of masses. Currently, in the LYRA project there are only $6$ haloes published, all with comparable halo masses of $\sim 10^9$\Msun. The new suite presented in this paper extends this to $65$ smaller haloes from $\sim 10^7$ to $10^9$\Msun. This new sample resolves the transition from classical dwarfs to ultra faint systems, and includes haloes unable to form stars. Thus, this work represents by far the largest sample of systems at any comparable resolution.

We additionally consider variations to the assumed radiation field in the early universe before reionisation ($z > 7$), and its effect on the simulated dwarf galaxies. In particular, we consider two variations in the Lyman-Werner  (LW) background. LW photons, with energies $11 < E/\rm{eV} < 13.6$, play a key role in the photodissociation of molecular hydrogen through the two-step Solomon process \citep{Solomon_65, Stecher_67}. It is well established that the strength of the LW background (LWB) plays a critical role in the formation of the first galaxies where, without the presence of metals, $\rm{H_2}$ is a key cooling channel. Many works have studied the effect of the LWB on the formation of a galaxy's first metal poor stars, using both dedicated numerical simulations \citep[e.g.][]{Machacek_01,Wise_12, Skinner_20, Schauer_21} and (semi-)analytic approaches to relate the Lyman-Werner background to a critical halo mass for popIII star formation \citep[e.g.][]{Trenti_09,Lupi_21,Nebrin_23}. These works have focussed on galaxies at high redshift, with their simulations not being evolved through to the present day, and are instead only evolved to high redshift ($z \sim 9$--$6$). As such, the connection between these assumptions about the LWB in the early Universe and the properties of present day galaxies remains unclear. In this paper we demonstrate that the early LWB plays a critical role in which haloes are able to form stars and has a significant impact on the properties of ultra faint dwarfs. The expected number of ultra faint dwarfs is particularly sensitive to the choice in the assumed LWB, while more massive systems ($\gtrsim 10^5$\Msun) are broadly unaffected.

This paper is organised as follows. In Section~\ref{Sec:Methods} we discuss the technical and numerical details of the simulations, giving particular attention to the aspects of the model that have changed from the simulations already presented in \cite{LYRA_2}. In Section~\ref{Sec:result_1} we present initial results from the new simulation suite, focussing on the dwarf galaxy-halo connection in the two different LW models. In Section~\ref{Sec:results_2} we study the origin of observed properties in the stellar mass-halo mass relation and halo occupation fraction, showing when galaxies first form their stars, their star formation histories, and which haloes they form in. In Section~\ref{Sec:discussion} we discuss important implications of our results, and summarise our findings in Section~\ref{Sec:summary}.

\section{Simulation details \& Methods} \label{Sec:Methods}

Here we outline the key aspects of the new simulations presented, and describe the details of how we process their outputs.

\subsection{Halo finder, merger trees and mass definitions} \label{section:halo_finder}

Individual gravitationally bound haloes are identified using the \texttt{SUBFIND} algorithm (last described in \cite{Subfind}). Initially, particles are assigned to separate groups using the friend-of-friends (FoF) algorithm with a linking length of $0.2$ of the mean interparticle separation. The FoF algorithm is run primarily on the DM particles, with gas cells and star particles attached through their nearest DM particle. Individual bound subhaloes within a given FoF group are identified from saddle points within the density field, and simulation particles assigned to subhalo through an iterative unbinding procedure. The most massive subhalo within a FoF group is identified as the central, or host, while all other subhaloes are considered to be satellites.

As is common throughout the literature we define the total halo mass, and size, as a sphere enclosing a mean density equal to $200 \rho_{\rm{crit}}$. Such definitions are commonly referred to as the `virial' mass and radius. We denote this enclosed mass and radius as \Mhalo$\,$ and $R_{\rm{200c}}$, respectively. For the stellar mass of a halo we use the bound mass, as calculated by \texttt{SUBFIND}. In this work we only present results for central host haloes and galaxies.

Merger trees are generated using the \texttt{D-haloes} algorithm, using only the collisionless DM particles to track progenitors, and is based on the methods proposed in \cite{Srisawat_13} and \cite{Jiang_14}. The algorithm uses the most bound particles of a given subhalo to track its descendent. From this initial linking merger trees are constructed, taking into account haloes missing in the \texttt{SUBFIND} catalogues at a given snapshot that may be linked to later snapshots.

\texttt{D-haloes} identifies the main progenitor branch based on the number of most bound particles linked by the algorithm. In our analysis it was found that \texttt{D-haloes} identification of the main progenitor often leads to haloes being lost (i.e. identified as having no progenitor). As we are interested in the very first star formation events in a halo's history (which in our simulation typically happens $z>10$) we need to reliably track progenitors to very high redshift. To this end, we define the `branch mass' as the total mass in all progenitors for a given subhalo. In this work we then choose the main branch as the progenitor with the largest branch mass. This allows us to reliably track haloes to very high redshifts, typically to the very first snapshot after the initial conditions ($z \sim 40$).

In some of our analysis we use the merger trees to trace when a galaxy first forms it stars. The primary analysis for this identifies when the main progenitor first contains stars. However, we found that in some cases the main progenitor does not correspond to the branch where the first star forms. Instead, the stars are born in a different progenitor and accreted onto the main branch much later. We refer to these as `ghost galaxies' and identify them as any galaxy whose youngest star is older than $10^8$\rm{yrs} at the time when the main progenitor DM halo first has stars. For these, we track the oldest star in the halo today and identify the host halo at its birth, independent of the merger tree (diamond points).

The existence of these ghost galaxies has been proposed using semi-analytic and particle tagging methods in \cite{Ghost_galaxies}. The natural occurrence of such objects in cosmological simulations highlights the need to use the full merger tree for tracking the formation and evolution of galaxies in models based on DM only simulations. Our analysis suggests that these `ghost galaxies' will always occur when defining the main progenitor using only the DM halo mass. The only plausible way to prevent this would be to define the main progenitor using the stars themselves, which is only possible in hydrodynamic simulations.

\subsection{Sample selection \& zoom-in initial conditions}
\label{Section:sample_selection}

Throughout this work we make use of the zoom-in simulation technique \citep[e.g.][]{Bertschinger_01}. This method allows a spatially connected region of the initial conditions to have an increased resolution compared to the rest of the simulation volume. Allowing for the target region to be simulated at a significantly increased resolution that would be unaffordable for the whole volume, while still resolving the gravitational effects of nearby large scale structures. This technique can be used for both individual haloes \citep[e.g.][]{Aquarious} and larger regions \citep[e.g.][]{APOSTLE_1, COCO}. All zoom-in initial conditions for the simulations presented were generated with the \texttt{IC-Gen} code \citep{Jenkins_10,Jenkins_13}

In this work we aim to produce a sample of field haloes that cover a wide enough mass range to form classical dwarfs, ultra-faint systems and starless haloes. To be analogues to the field dwarfs observed in our own Universe we aim to simulate haloes in environments resembling the vicinity of our Local Group of galaxies. Below, we outline how candidate haloes were chosen.

We begin by choosing candidates from the Local Group analogues presented in the APOSTLE simulations \citep{APOSTLE_1, APOSTLE_2}, which themselves were originally identified from the DOVE dark matter only cosmological box \citep{Jenkins_13}. Both DOVE and APOSTLE use a now outdated set of cosmological parameters, which differ from the original LYRA simulations. For consistency with existing LYRA simulations, and further consistency with currently preferred cosmological parameters, we resimulate the APOSTLE volumes using the Planck 2013 $\Lambda$CDM parameters \citep{Planck_13}. Specifically, $\Omega_{\rm{m}} = 0.307$, $\Omega_{\Lambda} = 0.693$, $\Omega_{\rm{b}}=0.048$, $H_{0} = 67.77 \rm{km}\,\rm{s}^{-1}\rm{Mpc}^{-1}$, $\sigma_8 = 0.829$, $n_s = 0.96$. These zoom simulations select a $5.9\,\rm{Mpc}$ ($4\, h^{-1}\mathrm{Mpc}$) radius spherical volume for each APOSTLE Milky Way -- M31 halo pair, centred on their barycentre. These are simulated as collisionless (i.e. DM only) cosmological zooms using a high resolution particle mass of $m_{\rm{p}} = 1.33 \times 10^5$\Msun, such that $10^{7}$\Msun{} haloes are sampled with $\sim 100$ particles. Our halo sample are chosen from these Local Group simulations.

To be able to successfully perform a cosmological zoom simulation requires haloes to be relatively isolated, such that they do not become contaminated by low resolution particles. We choose candidate haloes that meet the two following isolation criteria: (i) the halo is not within the splashback radius of a larger halo, which we take to be $2.5 R_{\rm{200c}}$ \citep[e.g][]{Splashback}; (ii) The distribution of particles at the initial, Lagrangian, positions are relatively compact. We define the compactness, $\kappa$, as
\begin{equation}
    \kappa = \frac{M_{\rm{tot}}}{\Omega_{m,0} \,\rho_{\rm{c}} \cdot 4/3 \pi R_{\rm{min, Lag}}^3 },
\end{equation}
where $M_{\rm{tot}}$ represents the mass of particles selected at $z=0$ (here we use all particles within $2.5 R_{\rm{200c}}$), and $R_{\rm{min, Lag}}$ is the radius of the minimum bounding sphere of the selected particles in the initial Lagrangian space. $\Omega_{m,0}$ and $\rho_{\rm{c}}$ are the present day matter density and critical density, respectively. In this definition, $\kappa$ is dimensionless and in the range $0<\kappa<1$. $\kappa = 1$ represents an initial configuration that is perfectly spherical, while $\kappa \to 0$ corresponds to an increasingly disconnected and more diffusely distributed set of particles. In this work we apply a criteria of $\kappa > 10^{-2}$.

From the haloes that meet the above criteria we choose $260$, selected based on their present day halo mass. Initially, $130$  haloes were selected to be uniformly sampled in $\log \,$\Mhalo, in the range $M_{\rm{200c}} = 10^7$--$10^9$\Msun, with an additional $130$ sampled between range $M_{\rm{200c}} = 10^8$--$10^9$\Msun. This was to increase the statistics of the haloes around the transition from dark to luminous. These candidates were then resimulated to $z=0$ as collisionless (DM only) systems with a particle mass of $100$\Msun using the zoom-in technique. The high resolution region is defined by selecting all particles within a $5 R_{\rm{200c}}$ sphere at $z=0$ for each halo. We then confirm if each halo remains uncontaminated at the higher resolution, defining all haloes with no low resolution particles within $R_{\rm{200c}}$ as uncontaminated. From these we randomly chose $65$ systems to simulate with full hydrodynamics. In addition we resimulate $4$ haloes from the existing LYRA haloes.

Including all uncontaminated haloes within the high resolution region gives us a total sample size as follows. We resolve $\approx 150,000$ uncontaminated central haloes resolved with at least $100$ particles ($M_{\rm{200c}} > 8.9 \times 10^3 \rm{M_{\odot}}$), $90$ haloes with masses $10^7<$\Mhalo$/$\Msun$<10^8$, $39$ with $10^8<$\Mhalo$/$\Msun$<10^9$ and $3$ with $10^9<$\Mhalo$/$\Msun$<5 \times 10^9$. In Appendix~\ref{sec:suite_stats} we present the distribution of halo masses for the new simulation suite, including both the main haloes selected for each zoom simulation and all uncontaminated haloes. We also present tabulated statistics for the new suite, presenting halo and stellar properties.

\subsection{The LYRA galaxy formation model}

All hydrodynamic simulations presented in this paper use the LYRA galaxy formation model. A detailed discussion of the model can be found in \cite{LYRA_1} and \cite{LYRA_2}. Here we outline the key aspects of the model and identify changes from previously published simulations.

LYRA is based on the cosmological code AREPO \citep{AREOP_1, AREPO_2}. AREPO uses a flexible moving mesh discretisation scheme to solve the hydrodynamics equations, and a hybrid octree and particle mesh method for the gravity solver. The hydrodynamic simulations use a DM particle mass of $74.7$\Msun{} and a target gas cell mass of $4$\Msun. A minimum stellar particle mass of $4$\Msun{} is used. The simulations use a maximum physical softening length of $14.8\, \mathrm{pc}$ and $5.91\, \mathrm{pc}$ for DM and stars, respectively. Transition from comoving to physical softening lengths at $z=9$.

LYRA resolves a fully cold interstellar medium (ISM) down to $T = 10\,\rm{K}$, with an extremely high baryon target resolution of $4$\Msun. Such a high resolution, combined with dynamic and hierarchical time stepping, ensures that the cooling radii of individual supernovae are fully resolved. This resolves the `overcooling problem' without the need for the implementation of subgrid feedback models.

Star formation follows a Schmidt relation with an efficiency parameter of $\epsilon = 0.02$. Gas cells with $n_{\rm{H}} > 10^3 \rm{cm}^{-3}$ and $T<100\rm{K}$ are allowed to form stars, with the efficiency set to $\epsilon = 1$ for $n_{\rm{H}} > 10^4 \rm{cm}^{-3}$ \citep[see][for details]{LYRA_1}. Stars are stochastically sampled from a Kroupa initial mass function (IMF) \citep{Kroupa_01}. Here a minimum mass of $4$\Msun{} is used, where stellar particles more massive than this represent individual stars, while particles at $4$\Msun{} are modelled as simple stellar systems consisting of low mass stars. Massive stars ($>8$\Msun) assume the stellar lifetimes from \cite{Portinari_98},  with the energy, mass and metallicity returns based on \cite{Sukhbold_16}. For less massive asymptotic giant branch stars we take the lifetimes and metal returns from \cite{Karakas}.

\subsubsection{Cooling tables \& background radiation fields} \label{section:cooling_tables}

\begin{figure*}
    \centering
    \includegraphics[width=\linewidth]{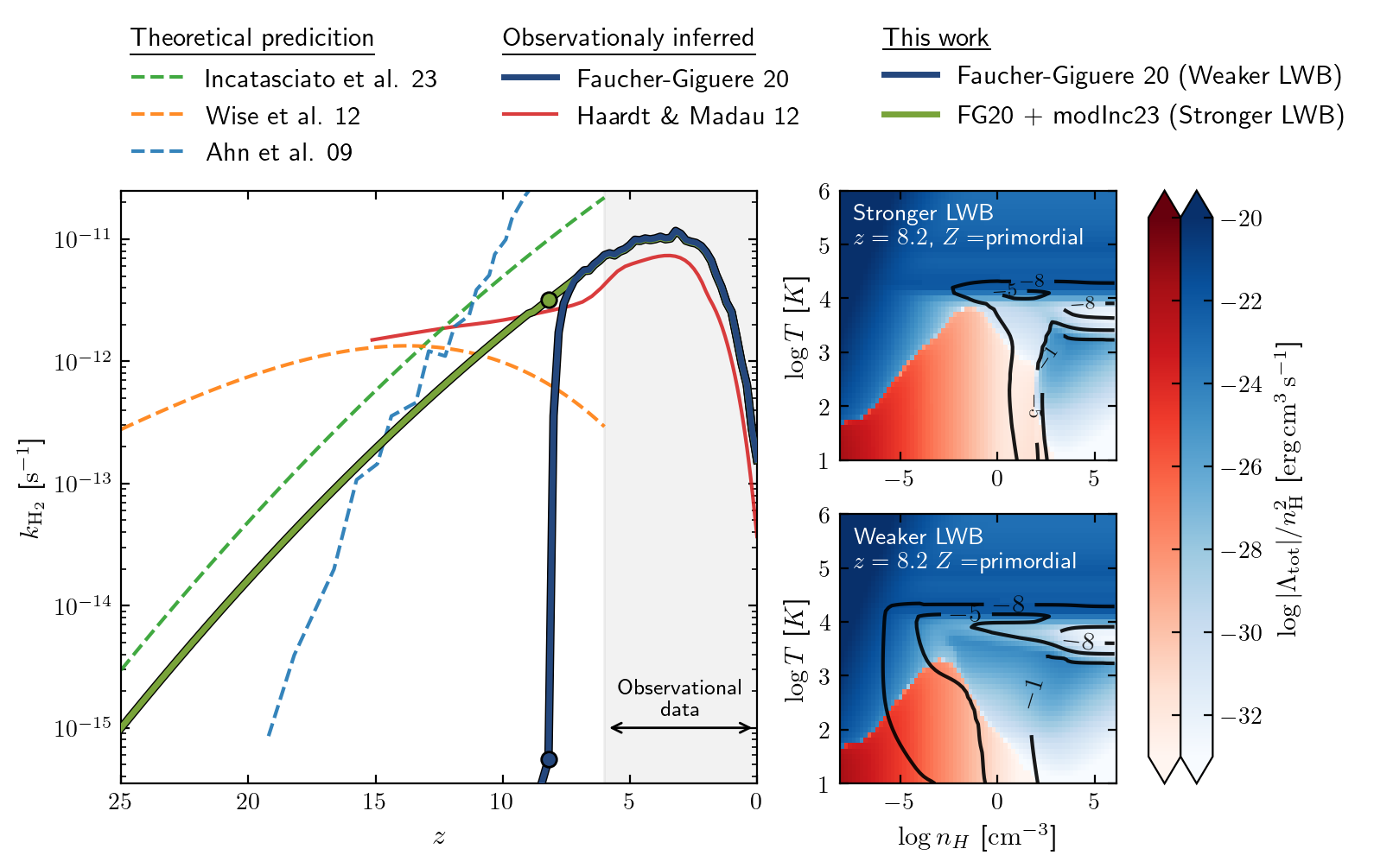}
    \caption{The left panel shows the redshift evolution of the LWB radiation ($E \approx 11$--$13.6\,\rm{eV}$), quantified through the $\rm{H_2}$ photodissociation rate, $k_{\rm{H_2}}$. In this work we consider two different backgrounds: one using the spectra from \protect\cite{FG20} (FG20) that has a negligible LWB prior to reionisation, while the second modifies the FG20 LW amplitude at high redshift to follow the functional form of \protect\cite{Inc23}, adjusted to match the amplitude of FG20 at $z=7$ (FG20 + modInc23). Throughout the rest of the paper we refer to these two spectra simply as the stronger and weaker LWBs, respectively. The two spectra only differ in the LWB for $z \gtrsim 7$, and share identical ionising spectral intensities (see Section~\ref{section:cooling_tables} for details). For comparison, we show a number of LWBs from the literature. This includes observationally-inferred backgrounds that are constrained using lower redshift data ($z \lesssim 6$) and extrapolated to higher redshifts \citep{Haardt&Madau, FG20}, along with a number of theoretical predictions from (semi-)analytic models and cosmological simulations \citep{Ahn_09, Wise_12,Inc23}. There is little consensus in the evolution of the LW at redshifts higher than $z \sim 10$ with significant differences between the models. In the right panels we show the net cooling rates from \protect\cite{Ploeckinger_25} at $z=8.2$ for the two LWBs used in this work. Gas in net cooling is shown in the blue colourmap and net heating in red. Overplotted are contours of constant $\rm{H_2}$ mass fraction, and the labels denote $\log\, 2 n_{\mathrm{H_2}}/n_{\mathrm{H}}$. At this redshift, the cooling function is strongly affected, with net cooling occurring at significantly higher densities for cool gas ($T = 10^3$--$10^4$K) in the stronger LWB.}
    \label{fig:LW_background}
\end{figure*}

A key aspect of any galaxy formation model is the heating and cooling of gas in the presence of a diverse range of radiation fields, temperatures, densities and chemical compositions. We model the contribution of H and He cooling using on-the-fly non-equilibrium ionisation rates from AREPO \citep{AREOP_1}, with the self-shielding model from \cite{Rahmati_13}. The contribution from molecular hydrogen, $\rm{H_2}$, and metals are taken from pre-computed cooling tables assuming equilibrium ionisation rates. For this, we use the tables from \cite{Ploeckinger_25}, which tabulate the net cooling and heating rates over a wide range of densities, temperature, metallicities and redshifts.\footnotemark The cooling rates are calculated using CHIMES \citep{CHIMES_1, CHIMES_2}, assuming equilibrium ionisation states. The $\mathrm{H_2}$ self-shielding implementation uses the fitting functions from \cite{CHIMES_2}, which reproduce the prediction from \textsc{CLOUDY} \citep{Cloudy}. The cooling tables model the contribution from $11$ elements and $20$ molecules. We refer the reader to the above references for further details.

\footnotetext{Previously published simulations \citep{LYRA_1, LYRA_2} use the cooling tables from \cite{Ploeckinger_20}, generated using CLOUDY \citep{Cloudy}. All simulations presented in this work use the updated tables from \cite{Ploeckinger_25}, generated with CHIMES \citep{CHIMES_1}.}

A key input needed to calculate the cooling rates is the assumed background radiation field, which sets the redshift evolution of the cooling tables. In this work we consider two different background radiation fields. Previously published LYRA simulations use the background radiation field from \cite{FG20} (from now on FG20). We begin with this as our fiducial starting spectrum, and consider a modification to the Lyman-Werner part of the spectrum at high redshift, as we discuss below.

This work focuses on the formation and evolution of small, metal poor, dwarf galaxies and the conditions in which they first form stars. In the absence of metals, $\rm{H_2}$ is the key cooling channel for primordial gas in the cold phase. $\rm{H_2}$ is readily photodissociated and destroyed by Lyman-Werner (LW) photons ($E \approx 11$--$13.6\rm{eV}$). In the early universe, before reionisation, ionising photons ($E > 13.6 \rm{eV}$) are readily absorbed by the intergalactic medium, however the Universe is mostly transparent to LW photons. As such, we expect a time varying, but spatially homogenous, LWB in the early Universe that plays an important role in the formation of the first metal-poor stars and galaxies.

In the left panel of Fig.~\ref{fig:LW_background} we present the $\rm{H_2}$ photodissociation rate, $k_{\rm{H_2}}$, as a function of redshift for the FG20 background alongside a number of others from the literature (see legend). It is common to also express the strength of the LWB through the $J_{21}$ factor, defined as $J_{\rm{21}} \equiv k_{\rm{H_2}}/(1.328 \times 10^{-12} \rm{s}^{-1})$. $J_{\rm{21}}$ can be interpreted as the intensity of the radiation for a given photodissociation rate, assuming low density gas in the optically-thin limit and a flat incident spectrum, in units of $10^{-21}\rm{erg}\, \rm{s}^{-1}\, \rm{Hz}^{-1} \, \rm{sr}^{-1} \, \rm{cm}^{-2}$ \citep{Abel_97}.

FG20 and \cite{Haardt&Madau} are observationally constrained background radiation fields, and commonly used throughout the literature. Both works provide tabulated spectra over a wide range of wavelengths and redshifts. In Fig.~\ref{fig:LW_background}, $k_{\rm{H_2}}$ is calculated directly from the provided spectra using CHIMES for these two works. \cite{Wise_12}, \cite{Inc23} and \cite{Ahn_09} are all predications from theoretical models for the high redshift ($z \gtrsim 8$) LWB. \cite{Wise_12} and \cite{Inc23} are taken from the fitting functions they present, while the \cite{Ahn_09} relation is taken directly from their plots.

At lower redshifts ($z <6$) the FG20 and \cite{Haardt&Madau} are in broad agreement, with similar amplitudes and functional forms for the LWB, agreeing within a factor of $\approx 2$. However, at higher redshifts ($z>7$), before reionisation, there is significant disagreement in the amplitude and redshift dependence of the LW radiation intensity. The theoretical predictions (dashed lines) differ by many orders of magnitude at $z \sim 30$--$15$, reaching some agreement, with an order of magnitude, at $z \sim 10$, before diverging again to lower redshifts. Even within the general disagreement, the FG20 spectra are unique in their redshift evolution, exhibiting a sharp decline in the amplitude at $z \sim 7$ not predicted by other relations. Furthermore, the LW part of the spectrum is essentially unconstrained at these redshifts in both FG20 and \cite{Haardt&Madau}, and instead represents an extrapolation of their respective fitting formulas.

To understand the impact of these differences in the early LWB radiation field, we run our simulations with two different assumptions about the LWB. The first is to use FG20 (Fig.~\ref{fig:LW_background} blue solid lines), which effectively represents the limiting choice that there is negligible LW radiation prior to reionisation. For the second spectra, we take the functional form of the redshift dependence from \cite{Inc23} and modify the amplitude to match FG20 at $z=7$. For $z>7$ we use
\begin{equation}
    \log J_{\rm{21}} = A + B(1+z) + C(1+z)^2,
\end{equation}
with $A = 1.642$, $B = -1.117 \times 10^{-1}$, $C = -2.782 \times 10^{-3}$, while for $z<7$ we directly take the FG20 spectra.

The two final background radiation fields only differ in the LW component at high redshift. For $z<7$, they are identical, while for $z>7$ the LW part of the spectrum is modified to obtain the $\rm{H_2}$ photodissociation rates shown in Fig.~\ref{fig:LW_background}. Importantly, the ionising part of the spectrum ($E>13.6\, \rm{eV}$) is identical over all redshifts, with the same assumed photoionisation and photoheating rates for $\rm{HI}$, $\rm{HeI}$ and $\rm{HeII}$ from FG20. In the FG20 model the hydrogen ionisation fraction is $0.5$ by $z = 7.8$. For simplicity we take this to be when reionisation occurs and denote this redshift as $z_{\mathrm{rei}}$ throughout the paper.

The primary effect of these two different UV backgrounds is on the net cooling rates in the simulations at these high redshifts. In the right two panels of Fig.~\ref{fig:LW_background}, we show the net cooling rates as a function of gas density and temperature. This is for the two assumed backgrounds at $z = 8.2$, chosen to be when the two differ the most, and at primordial metallicity. The colour map shows the normalised net cooling/heating rate, with blue for net cooling and red for net heating.

Focusing initially on the cooling rates from the weaker LWB (FG20, bottom right panel) a significant fraction of the phase space is in net cooling, with all gas $T > 10^{3.5} \rm{K}$ able to cool. The equilibrium line, where the heating and cooling terms are equal and where the colourmap changes from red to blue, has a peak temperature at relatively low densities, $n_{\rm{H}} \sim 10^{-3} \rm{cm^{-3}}$, and somewhat cool, $T \sim 10^{3.5} \rm{K}$. Gas that is able to be compressed to this density can undergo runaway cooling and form stars. For the stronger LWB (Inc23+FG20, top right panel), we see a similar structure to the cooling function, but the peak of the thermal equilibrium line is moved to higher densities, $n_{\rm{H}} \sim 10^{-1} \,\rm{cm^{-3}}$, and somewhat warmer gas, $T \sim 10^{4} \,\rm{K}$. Therefore, without external sources of gas compression, a much deeper gravitational potential is required to allow gas to reach the required density threshold for runaway cooling and collapse and lead to star formation. As we will show later, these differences in the cooling functions lead to significant differences in the halo mass at which star formation first occurs.

For reference, we additionally plot contours of constant $\rm{H_{2}}$ mass fraction in the right panels of Fig.~\ref{fig:LW_background} (solid black lines). Here, one can explicitly see the effects of the LWB. For the weaker background (FG20) a non-negligible amount of molecular hydrogen is able to form across a significant amount of the phase space. In contrast, the stronger LWB, $\rm{H_{2}}$ is restricted to denser gas. Above $n_{\rm{H}}\sim 10^2 \,\rm{cm^{-3}}$ molecular hydrogen self-shielding becomes important, allowing the gas to efficiently form $\mathrm{H_2}$ even in the presence of high intensity LW radiation. Above this density the cooling functions in the different LWBs are indistinguishable. The cooling tables used here all exhibit high $\rm{H_{2}}$ fractions and net cooling for $n_{\rm{H}} \gtrsim 10^2 \,\rm{cm^{-3}}$ and $T \lesssim 10^4 \,\rm{K}$ across all metallicities and redshifts.

Throughout the rest of the paper we refer to the FG20 background spectra and the one chosen to follow the modified \cite{Inc23} LW intensity relation as simply the weaker and stronger LWBs, respectively. Though, we re-emphasise that these spectra only differ for $z>7$.

\section{The dwarf galaxy halo connection} \label{Sec:result_1}

Here we present an initial analysis of the new suite of simulations, comparing how the two LW radiation fields in the early Universe ($z\gtrsim7$) affect the present day stellar content of dwarf galaxies and the connection to their host DM haloes.

\subsection{The stellar mass-halo mass relation} \label{sec:SMHM}
\begin{figure*}
    \centering
    \includegraphics[width=\linewidth]{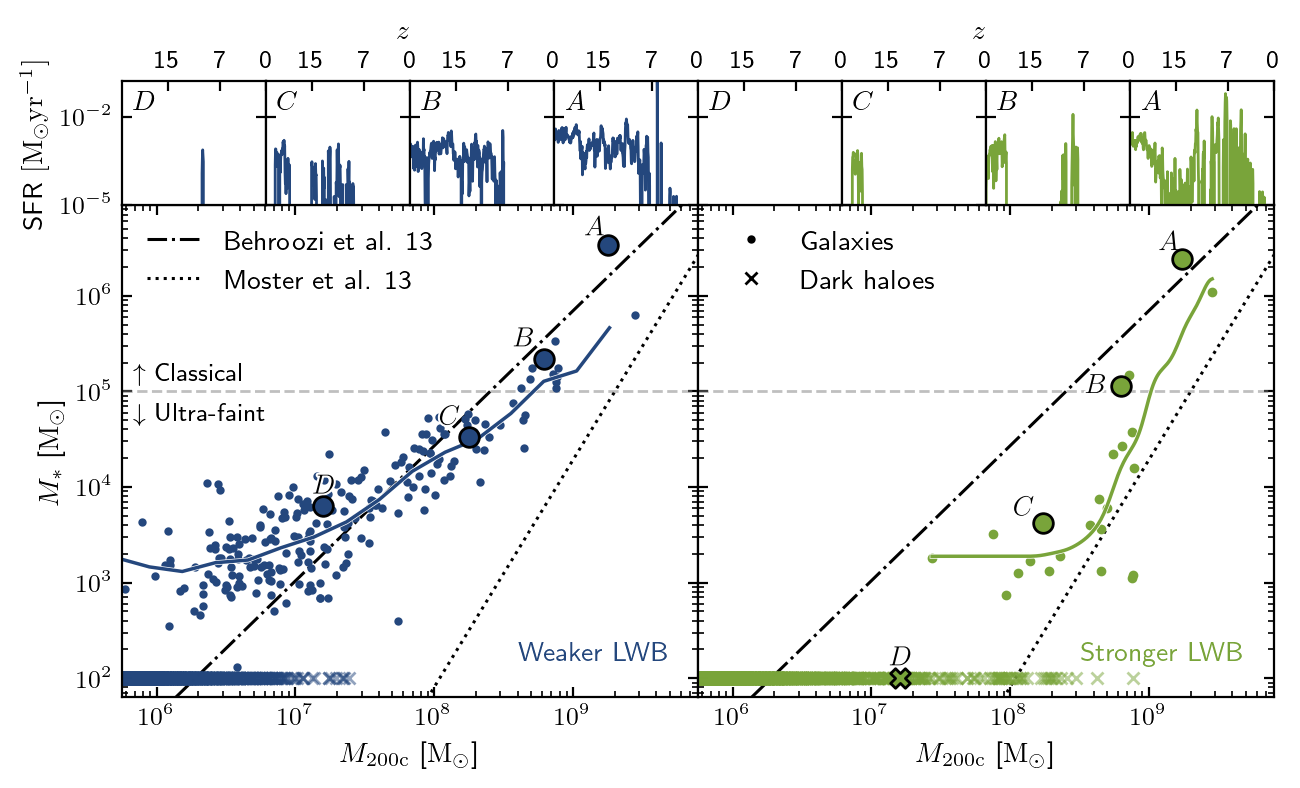}
    \caption{The present day ($z=0$) stellar mass-halo mass (SMHM) relation from our simulations for the two different LWBs. All central haloes that form stars are shown as circular points, with dark haloes that do not form any stars shown as crosses, and plotted at $M_{\ast} = 100$\Msun. We additionally plot the running median line for the luminous haloes as a solid line. For reference, we plot the extrapolated SMHM relation from \protect\cite{Behroozi_13} and \protect\cite{Moster_13}. The different LWBs have a significant impact on low mass dwarfs ($M_{\ast} \lesssim 10^6$\Msun) and which haloes are able to form stars. In the top panels, we show the star formation histories for four haloes (A, B, C, D, which are shown with larger symbols and a black outline), which encompass the range of observed formation histories. We use these haloes as case studies throughout the paper.}
    \label{fig:SMHM}
\end{figure*}

An insightful way of showing the diversity of galaxies in the new suite and the effect of the LW radiation on the dwarf galaxy population is through the stellar mass-halo mass (SMHM) relation. In Fig.~\ref{fig:SMHM} we show the present day ($z=0$) SMHM relation for all uncontaminated haloes within the new suite. Luminous galaxies are shown with circles, while dark haloes are plotted as crosses at $M_{\ast} = 10^2 \rm{M_{\odot}}$. Here we take `dark haloes' to be those without any bound stellar particles, as identified by \textsc{subfind}. The larger circles with black outlines highlight four `case study' haloes that we will study in detail throughout the paper.

Additionally shown for reference as dot-dash and dotted lines, respectively, are the SMHM relations from two popular abundance matching relations, \cite{Behroozi_13} and \cite{Moster_13}. We note that neither relation is constrained over the halo mass range presented here, and the curves represent extrapolations of their fits from more massive systems (the minimum constrained halo mass is \Mhalo$\sim 10^{10} \rm{M_{\odot}}$).

Considering the sample as a whole, we probe galaxies from small ultra-faints, with the smallest system being $M_{*} \approx 200$\Msun, through to more classical dwarf galaxy masses with $M_{*} \sim 10^6$\Msun. We also probe a very large range of halo masses, with the most massive halo in our sample being $M_{\rm{200c}} = 5 \times 10^9 \rm{M_{\odot}}$, while haloes of mass $M_{\rm{200c}} >7 \times 10^3  \rm{M_{\odot}}$ are resolved with more than $100$ DM particles. The range of halo masses shown in Fig.~\ref{fig:SMHM} covers all luminous haloes in our sample, with all haloes below this remaining dark today.

Focusing on the SMHM relation for the weaker LWB (left panel), we see that the transition from dark to luminous haloes occurs at $M_{\rm{200c}} \sim 10^6$--$10^7 \rm{M_{\odot}}$. In this mass range the SMHM relation for luminous haloes is relatively flat with near constant stellar mass at $M_{*} \sim 10^3 \rm{M_{\odot}}$. We emphasise that the baryon resolution, in both stars and gas, is $4 \, \rm{M_{\odot}}$, so this mass scale does not reflect the resolution limit of the simulations. At larger halo masses (\Mhalo$>10^7$\Msun) the relation steepens and becomes close to a power law with a relatively shallow slope ($M_{*} \propto M_{\rm{200c}}^{1.2}$). Interestingly, this change in the slope of the SMHM relation occurs at the same mass scale where haloes begin to transition from dark- to luminous. The SMHM relations are relatively smooth and with a near constant halo-to-halo scatter, with a standard deviation of $\approx 0.35 \, \rm{dex}$.

For the stronger LWB (right panel) we see a significantly different behaviour to the SMHM relation. One of the most notable differences is the minimum halo mass where haloes are able to host a galaxy, now occurring at \Mhalo$\sim 10^8$\Msun. Here, the increased LWB at early times prevents many of the smallest haloes from ever being able to form stars. Interestingly, both models exhibit a clear minimum stellar mass at $M_{*} \sim 10^3$\Msun, though this feature occurs at different halo masses. The overall form of the SMHM is significantly different between the two LWBs; while the weaker LWB is a relatively smooth relation, the stronger LWB has a sharp transition at \Mhalo$\approx 5\times 10^8$\Msun. Below this mass the SMHM is near constant, while it dramatically steepens above this halo mass. At \Mhalo$\sim 10^9$\Msun we find dark haloes and galaxies that differ by two orders of magnitude in stellar mass being hosted by comparable mass haloes. For haloes more massive than \Mhalo$\sim 5\times10^8$\Msun, the SMHM relation roughly follows a power law with slope $M_{*} \propto M_{\rm{200c}}^{3}$. For the more massive galaxies ($M_{\mathrm{\ast}} \gtrsim10^5$\Msun) we find that the different LWBs only moderately effect the present day stellar mass, typically by a factor of $\approx 2$. Overall, the main differences in the SMHM relation between the two different LWBs occurs in the low mass, ultra-faint regime.

A number of works have noted the existence of a `dip' in the satellite luminosity function between, with galaxies $M_{\ast} \approx 10^4$--$10^5$\Msun{} being less abundant than their larger and smaller cousins \citep[e.g.][]{Newton_18,Bose_18, Milky_census}, suggesting the existence of two distinct populations. Our stronger LWB, with the steep SMHM relation at higher mass and floor going lower masses, seems to naturally produces these two distinct populations. While the weaker LWB, with its smooth and shallow SMHM relation, does not lead to two distinct populations. Previous works have shown that the origin of these two distributions is likely imprint of cosmic reionisation with the two population being those that are able to restart star formation post-reionisation, and those that remain quenched. In our work we find a similar origin in the stronger LWB, but in our weaker LWB haloes are allowed to form stars in much smaller haloes which effectively smooths out any distinction between the two populations.

A similar feature to the minimum stellar mass at $M_{\ast} \approx 10^3$\Msun, present in both LWBs, has been observed in other theoretical works. Notably \cite{Wheeler}, using the FIRE-2 simulations at a resolution $m_{\rm{p}} \approx 40$\Msun, find a similar floor to the SMHM at $M_{\ast} \sim 10^3$\Msun. In addition, a number of SAMs predict similar behaviour. The A-Sloth model \citep[][]{A-sloth} predicts a clear floor, corresponding to $M_{\ast} \approx 10^4$\Msun. The GRUMPY \citep[][]{Grumpy_consttraint} and Galacticus \citep[][]{Galacticus_2} models predict a shallowing of the slope of the SMHM in the low mass end, but not a clear minimum stellar mass. In Section~\ref{Sec:results_2} we study the origin of this floor within our model, which, in short, is the consequence of these faint galaxies forming from a single burst of star formation at very early times in similar mass haloes.

Along with sampling a wide range of masses, we find a diversity of star formation histories. In the top panels, we show the star formation (SF) history, calculated from the ages of all stars at $z=0$, for four typical haloes that cover the sampled halo masses. Throughout the rest of the paper we will use these systems as individual case studies, and refer to them as Halo A, B, C and D (ordered in halo mass).\footnote{Previous LYRA simulations also refer to systems as Halo A, B, C, D etc. \citep{LYRA_3,Sureda}. The naming convention in this paper is \textit{not} consistent between these other works.} Here, we see that there are a number of distinct SF histories. All of these haloes form their first stars before reionisation, and for the most massive halo as early as $z \approx 25$. At the lowest stellar masses we find that these galaxies typically form from only one or two distinct bursts of star formation, and are quenched before reionisation (e.g. Halo D in the stronger LW model, and Halo C in the weaker LWB). At intermediate masses, there is more continual star formation up to reionisation, but are then quenched due to the ionising background and remain quiescent today (e.g. Halo C \& B in the weaker LW model, and Halo B in the stronger LW model). At the high mass end, the galaxies are not strongly affected by reionisation with Halo A able to continue to form stars through the epoch of reionisation, but is now quiescent. We also observe that differences in the LWB have a significant impact on the SF for the smaller haloes, as expected from the SMHM relation. In the extreme case of Halo D, this changes whether the halo is able to form any stars or not. We discuss the detailed differences between these systems later in Section~\ref{Sec:SF_histories}, and show here the SF histories to highlight the diversity observed in the new simulation suite.

\subsection{Halo occupation fraction}

\begin{figure*}
    \centering
    \includegraphics[width=\linewidth]{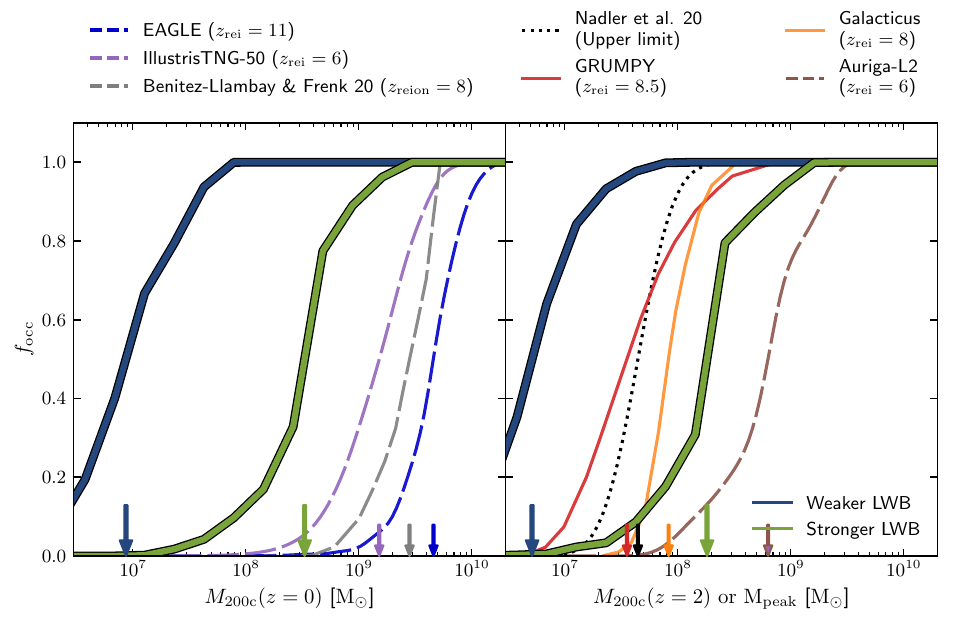}
    \caption{The halo occupation fraction, defined as the fraction of haloes containing any stars, for central haloes in the two LWBs. For reference, we plot a number of predictions from the literature. The arrows denote the mass where each model predicts $f_{\rm{occ}} = 0.5$. In the left panel we plot models for field haloes today, including the EAGLE and Illustris-TNG50 cosmological simulations and the (semi-)analytic model of \protect\cite{BL_20}. In the right hand panel we show works modelling Milky Way satellites, which use a satellite's peak halo mass before infall. Additionally in the right panel we show our $z=2$ results for comparison, a typical accretion time for Milky Way satellites. We show the prediction from the semi-analytic models GRUMPY and Galacticus and the purely empirical constraints from \protect\cite{Nadler_20} and the Auriga L2 resolution simulations. The line style denotes which cooling channels are considered in these works; solid lines are works that allow for molecular hydrogen cooling while dashed lines are works with only atomic cooling considered. There is a clear dichotomy, with models that only consider atomic cooling predicting the transition from dark to luminous haloes at \Mhalo$\sim 10^{9}$\Msun, while those modelling H$_2$ cooling and a cold ISM predict \Mhalo$\lesssim 10^{8}$\Msun. In the absence of a significant LWB prior to reionisation, the transition can be as low as \Mhalo$\sim 10^{7}$\Msun.}
    \label{fig:Occupation_fraction}
\end{figure*}

As we have seen, the different LWBs have a strong impact on which haloes are able to form stars. A common way of quantifying this transition is through the fraction of luminous haloes as a function of halo mass, often referred to as the halo occupation fraction, $f_{\rm{occ}}$. In the left panel of Fig.~\ref{fig:Occupation_fraction} we show the present day occupation fraction for our haloes in the two LWBs. Throughout the paper we will refer to the mass scale at which haloes transition from dark to luminous to be where $f_{\rm{occ}} = 0.5$. We show this mass scale as an arrow on the bottom x-axis.

Both LWBs exhibit a smooth transition from $f_{\rm{occ}} = 0$ to $1$, following the general form of a logistic curve. However, the transition from dark to luminous occurs on significantly different mass scales. In the weaker LW model, it is at \Mhalo$\sim 10^7$\Msun, with some galaxies hosted by haloes as small as \Mhalo$\sim 10^6$\Msun. For the stronger LWB, the transition occurs at \Mhalo$\sim 10^8$\Msun. It is clear that the choices in the LWB at high redshift can lead to significant changes to the present day occupation fraction, in this case a difference of more than an order of magnitude. In both models, the halo mass at which $f_{\mathrm{occ}} = 0.5$ is far below the present day atomic cooling limit, $M_{\rm{200c}} \sim 10^9$\Msun.

In Fig.~\ref{fig:Occupation_fraction} we also plot a number of predicted and inferred occupation fractions from the literature. In the left panel we show model predictions for the occupation fraction of central, field haloes. This includes results from two popular large volume cosmological hydrodynamic simulations, EAGLE \citep[using data from the L025N526 recal simulation][]{EAGLE_1, EAGLE_2} and Illustris-TNG50 \citep{TNG1, TNG2}, which have baryon resolutions of $m_{\rm{p}} = 2.3 \times 10^5$\Msun and $8.5 \times 10^4$\Msun, respectively. We also present the predictions from \cite{BL_20}, who developed a semi-analytic model for how the occupation fraction varies as a function of when reionisation occurs. Here we take their occupation fraction for $z_{\rm{reion}} = 8$, consistent with what is assumed in our simulations.

In the right hand panel of Fig.~\ref{fig:Occupation_fraction} we present works modelling the satellite population of the Milky Way, or Milky Way analogues. For our simulations we present the $z=2$ occupations fraction, a typical accretion redshift for Milky Way satellites. For the works from the literature the peak mass before infall is used. We present results from the Auriga L2 resolution Milky Way zoom simulation \citep{Auriga_L2}, which is one of the highest resolution Milky Way mass zooms with full hydrodynamics run to date with a mass resolution of $m_{\rm{p}} = 8.5 \times 10^2$\Msun. We present three models that are constrained to match the reconstructed Milky Way satellite properties and counts \citep[][]{Sat_counts}; the semi-analytic models GRUMPY \citep{Grumpy_model, Grumpy_consttraint} and Galacticus \citep{Galacticus_1, Galacticus_2} and the empirical, upper limit, constraints from \cite{Nadler_20}.\footnote{The Galacticus model uses a different overdensity threshold for their mass definition compared to the other works, taken from \cite{Bryan_norman_98}. We do not apply any corrections to their occupation fraction, but note that typical differences between this definition and \Mhalo, used for all other works including our own, is $\sim 10$\%.}

The three models constrained by Milky Way satellites are broadly consistent with each other and sit between our two simulations, with the weaker LW being significantly below all models and the stronger LWB at a somewhat higher mass. These results are broadly consistent with the recent work of \cite{Nadler_25}, who similarly found that \cite{Inc23} LWB with FG20 ionisation history predicts an occupation fraction at slightly higher masses than the observationally inferred values suggest. It is clear that one could choose a LWB that reproduces the observationally inferred occupation fraction. However, one could also vary other aspects of the simulations other than the LWB to reproduce the occupation fraction, in particular the assumed reionisation redshift.

The different line styles in Fig.~\ref{fig:Occupation_fraction} denote which cooling channels are modelled. Dashed lines represent models that only consider atomic hydrogen cooling and do not model a cold ISM. Solid lines are used for works that explicitly model the cold phase and include $\rm{H_2}$ cooling, while the dotted line represents a versatile empirical approach that does not explicitly consider any cooling channels \citep{Nadler_20}. From Fig.~\ref{fig:Occupation_fraction}, a clear dichotomy arises where models only considering atomic hydrogen cooling predict the transition to dark haloes occurs at \Mhalo$\sim 10^9$, even for very high resolution simulations (e.g. Auriga-L2) or semi-analytic models with no resolution limit \citep[e.g.][]{BL_20}. Works that model the cold phase and $\rm{H_2}$ cooling allow for star formation in much smaller haloes, even with the same assumed reionisation redshift. All these models predict the transition from dark to luminous to be \Mhalo{}~$\lesssim 10^8$\Msun.

This highlights the importance of modelling the contribution from $\rm{H_2}$ to reproduce the full population of faint dwarf galaxies. If this key cooling channel is not considered, models are likely artificially restricting star formation to larger haloes.

\section{The formation and evolution of our dwarf galaxies} \label{Sec:results_2}

In the previous section we saw a number of key features in the SMHM relation and occupation fraction. Both models predict a clear minimum stellar mass at $M_{\ast} \sim 10^3$\Msun. The weaker LWB predicts that the transition from dark to luminous haloes occurs at \Mhalo$\sim 10^7$\Msun{}, as well as a smooth SMHM relation. On the other hand, the stronger LWB predicts the transition to occur at \Mhalo$\sim 10^8$\Msun{} with a sharp break in the SMHM relation at the same halo mass scale. In the next section, we explore how these similarities and differences come about, and the physical origin of the final present day galaxy-halo connection.

\subsection{Critical mass for star formation} \label{sec:critical_mass}

\begin{figure*}
    \centering
    \includegraphics[width=1.0\linewidth]{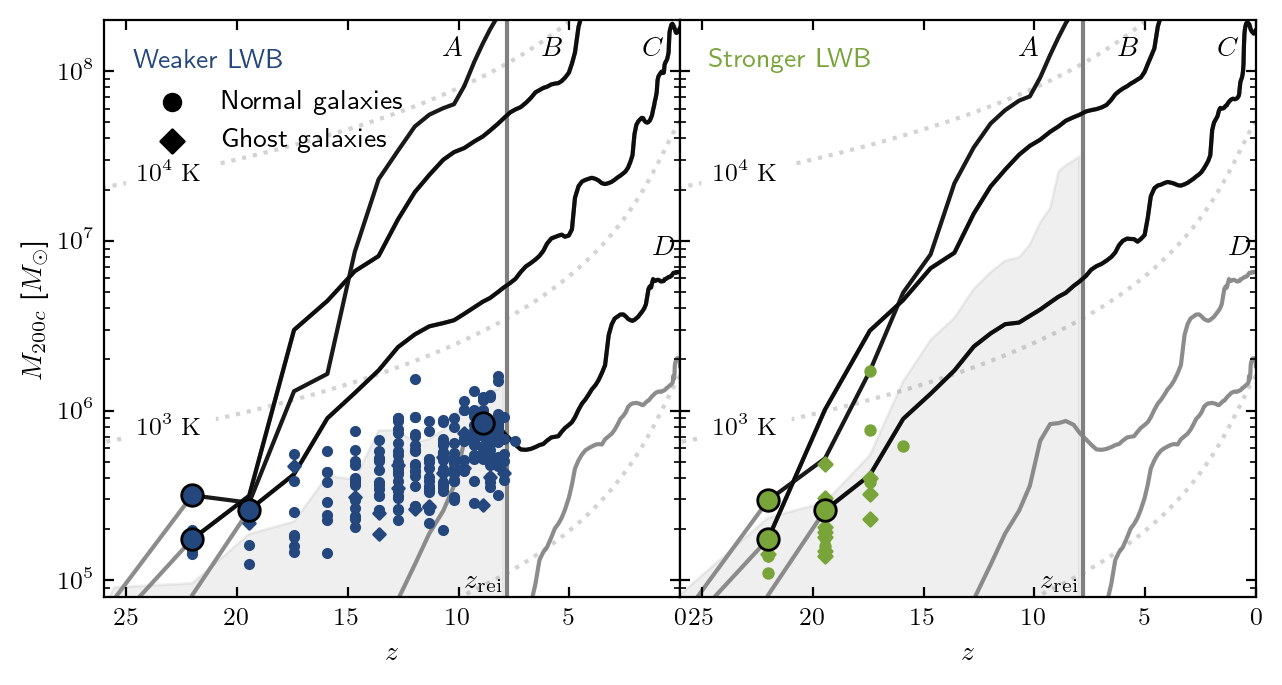}
    \caption{The halo mass and redshift at which each galaxy first forms stars. This is primarily calculated when the main progenitor first has stars (circular points). For some galaxies, their stars are not born in the main branch and instead accreted much later. For these systems the birth time of the oldest star and its host halo mass is identified, regardless of the merger tree -- we refer to such objects as `ghost' galaxies (diamond points). The shaded grey region corresponds to the largest halo that has not formed stars at each redshift. We plot the redshift of reionisation and lines of constant virial temperature, $T_{\rm{200c}}$, for reference. In the weaker LWB, SF occurs when haloes reach \Mhalo$\sim 10^{5.5}$\Msun up to reionisation, while haloes in the stronger background form stars at similar masses but are restricted to higher redshift ($z \gtrsim 15$) before the LWB becomes sufficiently strong. We plot the halo mass growth for the four case study haloes, along with a smaller halo that remains dark in both models. Grey lines are used when the halo is dark, and black after the halo has formed stars.}
    \label{fig:formation_mass}
\end{figure*}

We begin this analysis by studying when galaxies form their very first stars. In many theoretical models, including both analytic calculations and cosmological simulations, there exists a well defined, redshift dependent, critical mass threshold, where haloes that are able to cross this critical mass during their evolution are able to begin star formation.

In Fig.~\ref{fig:formation_mass} we show the redshift and halo mass when each galaxy in our sample first forms stars. For the majority of haloes this occurs the first star formation occurs in the main progenitor of the halo (circles), but for some this happens in a different halo progenitor, with these stars accreted much later than they are formed (diamonds). We refer to these as ghost galaxies; see Section~\ref{section:halo_finder} for a discussion of how these are identified and defined.

Considering both models together, we see a number of shared features. No galaxies in our sample form their first stars after reionisation. While we are limited in our statistics, it does suggest that for these LWBs the majority of galaxies first ignite star formation in the early Universe ($z \gtrsim 7 $). The halo mass at which star formation first occurs is remarkably similar for all haloes in both LWBs, in the range \Mhalo{} $\sim 10^5$ -- $10^6$\Msun, far below the atomic cooling limit ($T_{200c} = 10^4\rm{K}$) at any redshift.

For the weaker LWB (left panel) we see a clear critical mass threshold. This is centred on \Mhalo$\approx 10^{5.5}$\Msun, following a log normal distribution with a standard deviation of $\approx 0.2 \, \rm{dex}$. There is also a slight redshift dependence to the typical formation mass, consistent with a constant virial temperature of $T_{\rm{200c}} \sim 10^{2.5} \rm{K}$. The earliest star formation occurs at $z \sim 20$ with haloes beginning star formation up to $z_{\mathrm{rei}}$.

The key difference between the weaker and stronger LWB (right panel) is that haloes in our sample do not continue to undergo their first episode of star formation up to reionisation. Instead, this model does not have the population of galaxies that begin star formation at $z \sim 7$--$15$. Those haloes that do form stars in both LWBs do so at a comparable mass, with haloes at larger masses around $z \sim 15$. With our finite statistics, we identify that the criteria for star formation is if a halo can cross the \Mhalo{} $\sim 10^5$\Msun threshold early ($z \sim 20$) while the LWB is relatively weak. If crossing this same mass threshold later when the LWB is significantly stronger, the halo is unable to efficiently cool and begin star formation.

Based on other works in the literature, there is probably a well defined critical mass for star formation at $z \approx 15$--$7$ in the stronger LWB, likely at \Mhalo$\sim 10^7$\Msun \citep[e.g.][]{Skinner_20, Nebrin_23}. With the finite statistics of our suite this suggests that haloes forming their first stars at this time are relatively rare, with the majority of haloes having either crossed the \Mhalo{} $\sim 10^5$\Msun threshold earlier at $z\sim20$, or remain below \Mhalo{} $\sim 10^7$\Msun{} up to reionisation.

We additionally tested the critical threshold when expressed as a halo's maximum circular velocity, $V_{\rm{max}}$. When expressed this way, the galaxies typically begin forming stars at $V_{\rm{max}} \approx 3$--$\,4\,\rm{km}\,\rm{s}^{-1}$. We do not find that $V_{\rm{max}}$ meaningfully reduces the scatter in critical threshold, with a standard deviation of $\approx 0.2\,\rm{dex}$ for \Mhalo$\,$ and $\approx 0.1\,\rm{dex}$ for $V_{\rm{max}}$, broadly consistent with the same intrinsic scatter (recalling that $v_{\rm{circ}} \propto M^{1/2}$).

Fig.~\ref{fig:formation_mass} also shows the halo mass growth of the main progenitor for the four case study haloes, A, B, C \& D, along with a halo that is unable to from stars in either model and has a present day halo mass of \Mhalo$\sim 10^6$\Msun. The line colour denotes if the halo has yet formed stars (black) or not (grey). For halo A, B and C, the different LWBs have little to no effect on the halo mass and redshift of the first star formation, with any differences being due to slight changes in the main branch identification. In contrast, halo D goes from being a small ultra-faint galaxy that is able to form stars just prior to reionisation at $z \approx 9$ in the weak LWB, to being unable to form any stars in the stronger LWB model. The halo that is unable to form stars in either LWB, remains below the typical formation mass before reionisation and is only able to cross this threshold at $z \approx 5 $, far after reionisation.

\begin{figure}
    \centering
    \includegraphics[width=\linewidth]{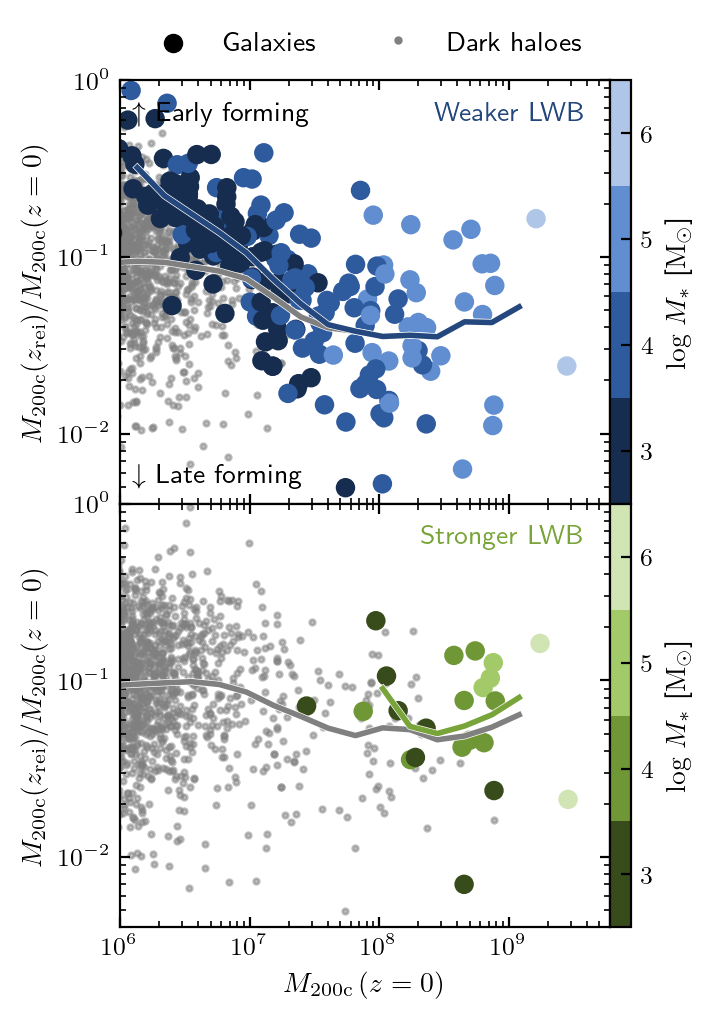}
    \caption{The halo mass growth since reionisation as a function of present day halo mass. The small grey points correspond to dark haloes at $z=0$, while the larger coloured points are haloes that host a galaxy, with the shade denoting the present day stellar mass (see colourbar). The top and bottom panels show the weaker and stronger LWBs, respectively. Overall at a fixed halo mass, earlier forming haloes are more likely to form stars and host a larger galaxy. However, this correlation is not as strong in the weaker LWB.}
    \label{fig:Bias_accretion}
\end{figure}

The present day occupation fraction (Fig.~\ref{fig:Occupation_fraction}) arises from the critical mass curve convolved with the range of halo accretion histories. In many models this results in a correlation between the stellar mass and halo formation time at a fixed present day halo mass, with earlier forming haloes more likely to be luminous and form more stars \citep[e.g.][]{Fitts_17, Rey_20, BL_20}. In Fig.~\ref{fig:Bias_accretion} we present the halo growth since reionisation, \Mhalo$(z=z_{\rm{reion}})/$\Mhalo$(z=0)$, as a function of present day halo mass. This is plotted for both dark and luminous haloes, with the colour correspondence to the galaxies present day stellar mass.

For the weaker LWB (top panel), all haloes above \mbox{\Mhalo$(z=0)\sim 10^8$\Msun{}} host a galaxy. Below this mass, luminous haloes are systemically those that form earlier than the overall population, with the bias increasing when moving to lower masses. For those systems that do form stars, there is a strong correlation in the stellar mass with accretion history. At a fixed halo mass, earlier forming haloes generally form more stars, consistent with results from the literature \citep{Fitts_17,Rey_20}.

For the stronger LWB (bottom panel) we do not see a strong correlation with formation time for a halo's ability to form stars. With the grey and coloured line closely following each other. For luminous haloes, there is a mild correlation in the total amount of stars formed. We have tried different parameterisation for the accretion history, varying the redshift at which we take the early halo mass from $z=20$ to $z=4$, along with studying the correlation with the peak circular velocity and halo concentration today, which are common proxies for earlier and later forming haloes. None of these reparametrisations exhibit a strong correlation with which haloes are able to form stars, or the amount of stellar mass formed. We note that almost all of the galaxies shown are quenched at reionisation; rejuvenating haloes in the LYRA model show a clear correlation between the present day stellar mass and formation time \citep{LYRA_3, Sureda}.

Interestingly, in both LWBs we find many haloes that remain dark despite having comparable masses at the redshifts when the luminous galaxies first form stars. This is most clearly shown in Fig.~\ref{fig:formation_mass}, where the grey shaded region shows the mass range where we observe dark, starless, haloes in our sample at each redshift. This region significantly overlaps with the population of haloes that are beginning star formation, suggesting that, while the mass accretion history is most probably the dominant factor, there are important secondary properties that lead to a halo beginning star formation.

\subsection{Gas phases at a galaxy's birth}

\begin{figure*}
    \centering
    \includegraphics[width = \textwidth]{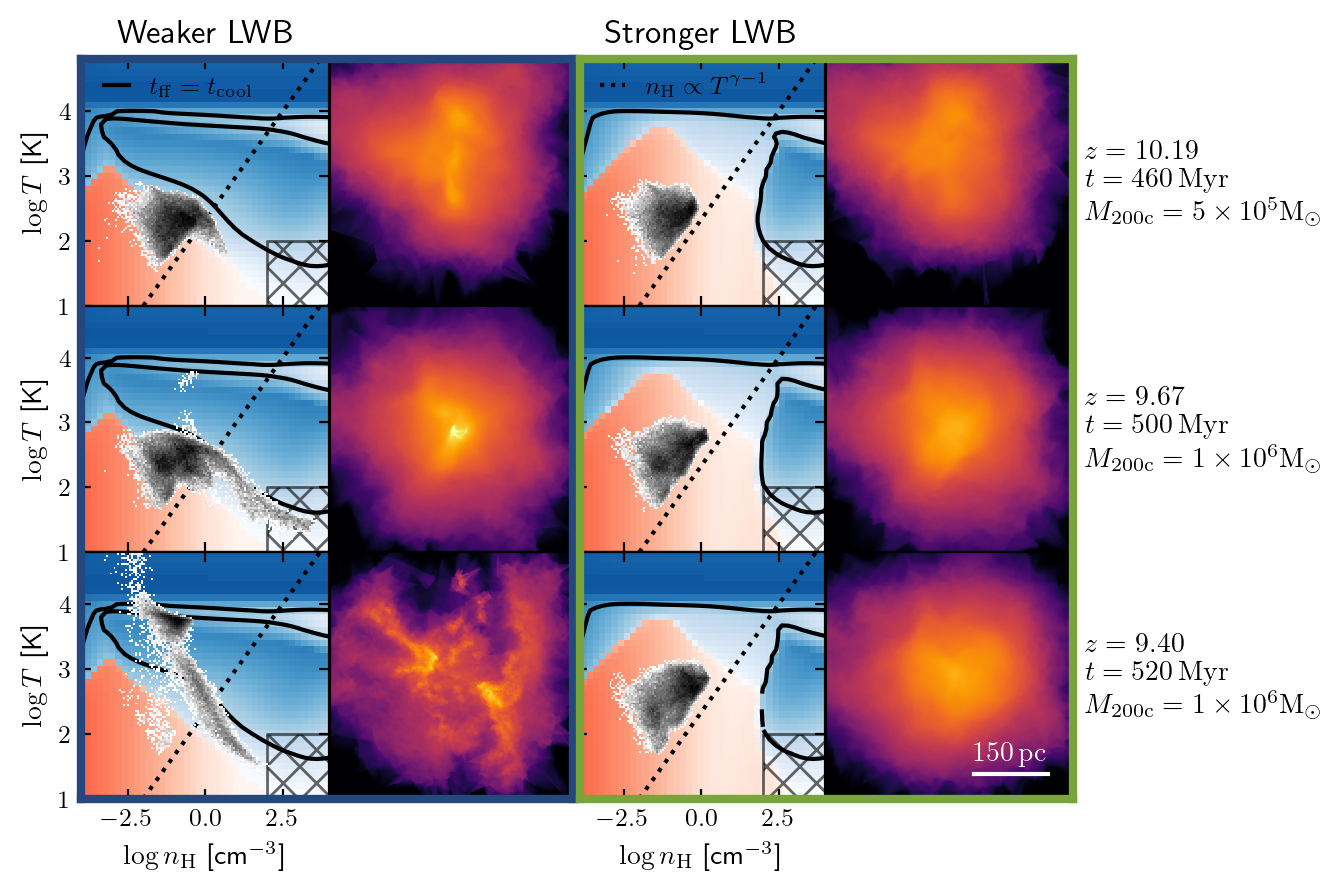}
    \caption{The spatial and phase space distribution of gas in Halo D. Time runs from top to bottom. The weaker and stronger LWBs are shown in the left and right panels, respectively. The particular times are chosen to be just before star formation (top), during star formation (middle), and just after the first SNe (bottom) in the weaker LWB; the halo is unable to form stars in the stronger LWB. In the phase diagram, the distribution of gas cells is shown with the grey points. The cooling function is shown for reference. The contour where the cooling time and free fall time are equal ($t_{\mathrm{ff}} = t_{\mathrm{cool}}$) is shown as the solid black line, an adiabatic relation plotted as the dashed black line ($n_{\mathrm{H}} \propto T^{\gamma - 1}$), the star formation criteria is plotted as the hashed box in the bottom right ($T<100\, \mathrm{K}$, $n_{\mathrm{H}} > 10^3\, \mathrm{cm^{-3}}$). The visualisations show the column density of the gas. On the right of the figure the halo mass, redshift, and time for each row are specified.}
    \label{fig:first_stars_halo_134}
\end{figure*}

In this section, we study the gas phases during the first epochs of star formation. We use Halo D as an example. This is an interesting case where in the weaker LWB it forms its stars through a single burst at $z\sim10$, while in the stronger LWB it is unable to form any stars at all (see Fig.~\ref{fig:SMHM}).

In Fig.~\ref{fig:first_stars_halo_134} we show the phase space of the gas (temperature, $T$, and density expressed as the hydrogen number density, $n_{\rm{H}}$) and a visualisation of the spatial distribution of gas. The left two panels show the system in the weaker LWB, and the right two show the stronger background. Each row shows the simulation at a particular time, chosen to be just prior to the first stars forming (top), during star formation (middle) and just after the first SN have occurred and are driving a strong outflow (bottom), in the weaker LWB. The phase space shows the distribution of gas cells in the grey colourmap, with the cooling function under-plotted, using the same normalisation as Fig.~\ref{fig:LW_background}.  

There are two key timescale to consider when investigating if gas is able to cool and collapse efficiently. The time it takes for gas to radiate away its internal energy, known as the cooling time
\begin{equation}
	t_{\mathrm{cool}} = \frac{k_{\mathrm{B}}}{\mu m_{\mathrm{p}}(1-\gamma)} \frac{T}{n_{\mathrm{H}} \Lambda (T, n_{\mathrm{H}})}.
\end{equation}
And the time for a gas cloud to collapse under its own gravity, known as the free fall time
\begin{equation}
	t_{\mathrm{ff}} = \sqrt{\frac{3\pi}{32 G \rho}}.
\end{equation}
In the above equations, $\mu$ denotes the mean molecular weight, for which we take values assuming neutral gas of primordial composition ($\mu = 1.21$, $X = 0.76$). We additionally take $\gamma = 5/3$, for mono-atomic gas.

If $t_{\mathrm{ff}} > t_{\mathrm{cool}}$ then gas is able to cool quicker than it can contract, leading to runaway collapse and subsequently star formation. While gas with $t_{\mathrm{ff}} < t_{\mathrm{cool}}$ cannot efficiently radiate away its internal energy and its collapse may be stopped by thermal pressure, reaching an equilibrium configuration. In practice the behaviour is more complex than these simplified equations, with the cooling function having a complex dependence on temperature and density, but it is still instructive to determine if gas is, instantaneously, in the regime of efficient cooling.

For additional reference we plot an adiabatic relation as a dashed black line ($\rho \propto T^{\gamma - 1}$). Gas that contracts or expands without heat transfer will move parallel to this line in phase space. Finally, in the bottom right of the phase space we show the star formation criteria used in the simulations as a hatched region. Only gas with $T<100\,\mathrm{K}$ and $n_{\mathrm{H}} > 10^3\, \mathrm{cm^{-3}}$ is able to form stars.

Focussing on the weaker LWB just before star formation occurs (top left panels), most of the gas occupies a relatively narrow range of the phase space, $T \sim 100\,$--$\,1000\,\rm{K}$, $n_{\rm{H}} \sim 0.01$--$\,1 \,\rm{cm^{-3}}$. This approximately straddles the equilibrium cooling line. A small fraction of the gas is beginning to cool, leading to the tail in the distribution to higher densities. However, all the gas is still in the $t_{\mathrm{ff}} < t_{\mathrm{cool}}$ regime. In the next snapshot (middle left panels) the gas has now been able to be compressed to the point where $t_{\mathrm{ff}} > t_{\mathrm{cool}}$ and is able to efficiently cool and collapse, with ongoing star formation at the centre of the halo. In the final snapshot presented (bottom left panels) the first SN have occurred, with the subsequent shocks driving a strong outflow that will quench the galaxy. We now see gas over a much wider range of temperatures and densities. Some gas has been heated to $T\sim 10^6 \, \rm{K}$, while the majority of the mass is relatively diffuse $n_{\rm{H}}\sim10^{-1}\, \rm{cm^{-3}}$ and just below the atomic cooling limit, $T\sim 10^4\, \rm{K}$. There remains some cold dense gas that is unable to cool efficiently and is confined within the outflow. Shortly after the final snapshot shown here, the halo is almost completely devoid of gas due to the SN-driven outflows. Halo D is then unable to (re)accrete sufficient gas to restart star formation before reionisation permanently quenches the system.

Focussing on the stronger LWB (right columns) we see significantly different behaviour. Initially, the gas occupies a similar part of the phase space as seen in the weaker LWB, but without the tail of gas that is beginning to cool. The behaviour can easily be understood by the cooling function. While the gas is able to contract to the same densities, with the stronger LWB it is more difficult to form $\rm{H_2}$ and this part of the phase space is no longer in net cooling. When looking at the later snapshots, we see that the gas does continue to moderately contract as the halo grows, roughly parallel to the adiabatic line. However, it is never able to reach densities where efficient cooling can begin ($\gtrsim 10^2\,\rm{cm^{-3}}$).

In Appendix~\ref{section:appendic_gas} we also present a similar analysis for Halo C. This halo differs from halo D by forming stars in both LWBs, and forming its first stars significantly earlier ($z \sim 20$). The first episode of star formation occurs in nearly identical ways for Halo C in the two LWBs, with only modest changes to the phase space distribution of its gas.

\subsection{Star formation histories} \label{Sec:SF_histories}

\begin{figure*}
    \centering
    \includegraphics[width=\linewidth]{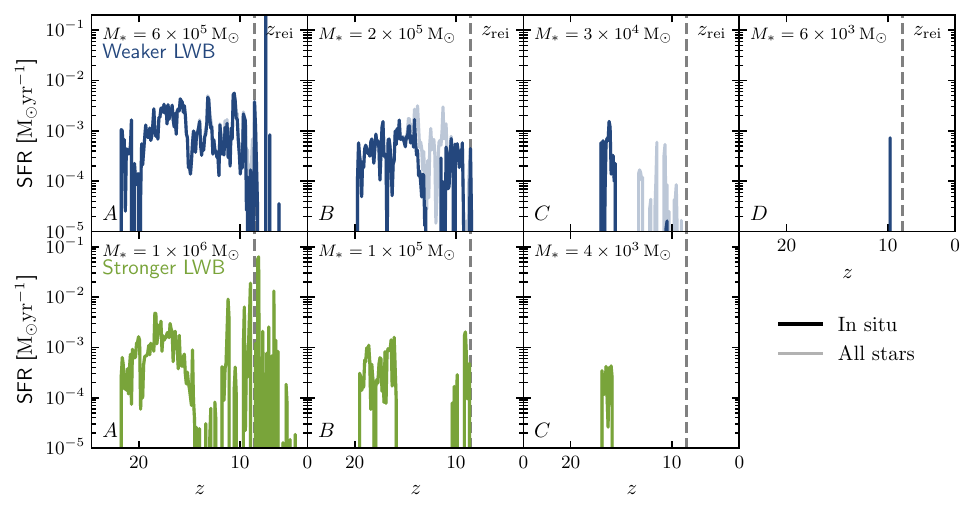}
    
    \caption{The star formation histories for Halo A, B, C \& D. The weaker LWB is shown in the top panels, and the stronger background on the bottom panels. In situ star formation, defined as all stars either born in situ to the main branch or accreted before reionisation, is shown with the solid line, while the transparent line is for all stars in the halo at $z=0$. Note that halo D in the stronger LWB does not form stars. We see a wide diversity of formation histories, from single star bursts at high redshift to systems that are able to continue star formation through reionisation.}
    \label{fig:SF_hist_individual}
\end{figure*}

In Fig.~\ref{fig:SF_hist_individual} we show the star formation (SF) histories for Halo A, B, C and D that sample the range of observed SF histories in our sample. The SF histories are calculated from the ages of all bound stars today, additionally split by those identified as in situ. In this work, we define `in situ' as stars identified as being born in the main progenitor or accreted before reionisation. As identifying, and following, the main progenitor beyond reionisation is not trivial, we apply this criteria such that we identify stars that are accreted onto the halo far after most of our sample have ceased star forming. This choice represents a conservative definition for the accreted stellar fractions.

The different LWBs have a moderate effect in the total stellar mass formed in Halo A, with the halo forming around twice as many stars in the weaker LWB. The initial SF histories are comparable in the two models, first beginning star formation at $z \approx 22$ at comparable SF rates. In the stronger LWB the halo is able to continually form stars through reionisation, with moderate levels of star formation close to the present day. In the weaker LWB the halo temporarily ceases star forming at reionisation, before undergoing a strong burst of star formation at $z \approx 6$, and undergoing little star formation after this.

For halo B, we see that the first epochs of star formation are also broadly unaffected by the changes to the LWB. Star formation begins at the same time ($z \approx 20$) with similar SF rates. In the stronger LWB (bottom panel), the galaxy goes through a prolonged quenched phase before restarting SF just prior to reionisation. In the weaker LWB (top panel), the galaxy is able to form stars nearly continually until reionisation, and additionally grows through accretion.

Halo C shows similar in situ formation between the two LWBs. In the weaker LWB, it undergoes one distinct in situ SF episode at $z \sim 15$, with some moderate levels of SF just prior to reionisation. However, the galaxy continues to grow significantly through accretion, leading to a system dominated by accreted stars today, with only $36$\% of stars identified as in situ. In the stronger LW model, it is only able to undergo one burst of SF at high redshift ($z \approx 20$) and does not accrete any stars, with all smaller accreted haloes remaining dark in the stronger LWB.

Halo D has the most dramatic change to its star formation. In the weaker LWB it is able to form stars, just crossing the threshold for star formation prior to reionisation at $z \approx 10$. Here, similar to Halo C in the stronger LWB, it only forms one burst of stars and contains no accreted stars. In the presence of the stronger LWB, this halo is completely unable to form any stars, remaining dark until the present day.

We find that many of our fainter dwarfs form from a single episode of star formation. Traditionally, such a SF history is associated with stellar and globular clusters \citep[e.g.][]{Krumholz_19}. These objects, however, are distinct in their formation from typical star clusters, in particular forming within a host DM halo. Many such systems end up being accreted onto a larger galaxy. This can lead to their DM haloes being stripped, but retaining a bound stellar component that has properties consistent with globular clusters. This has been shown to happen in the more massive LYRA galaxies \citep{Gutcke_DM_clusters}. As such, today they would be observed as stellar populations from a single star burst at high redshift, be relatively compact, and DM free. The EDGE simulations have also recently reported such objects, which they refer to as globular-cluster-like dwarfs \citep{Taylor_25}. The formation of such objects appears to naturally arise in cosmological simulations that have sufficient resolution and which resolve the cold phase of the gas.

\subsubsection{SF timescales}

\begin{figure*}
    \centering
    \includegraphics[width=1.0\linewidth]{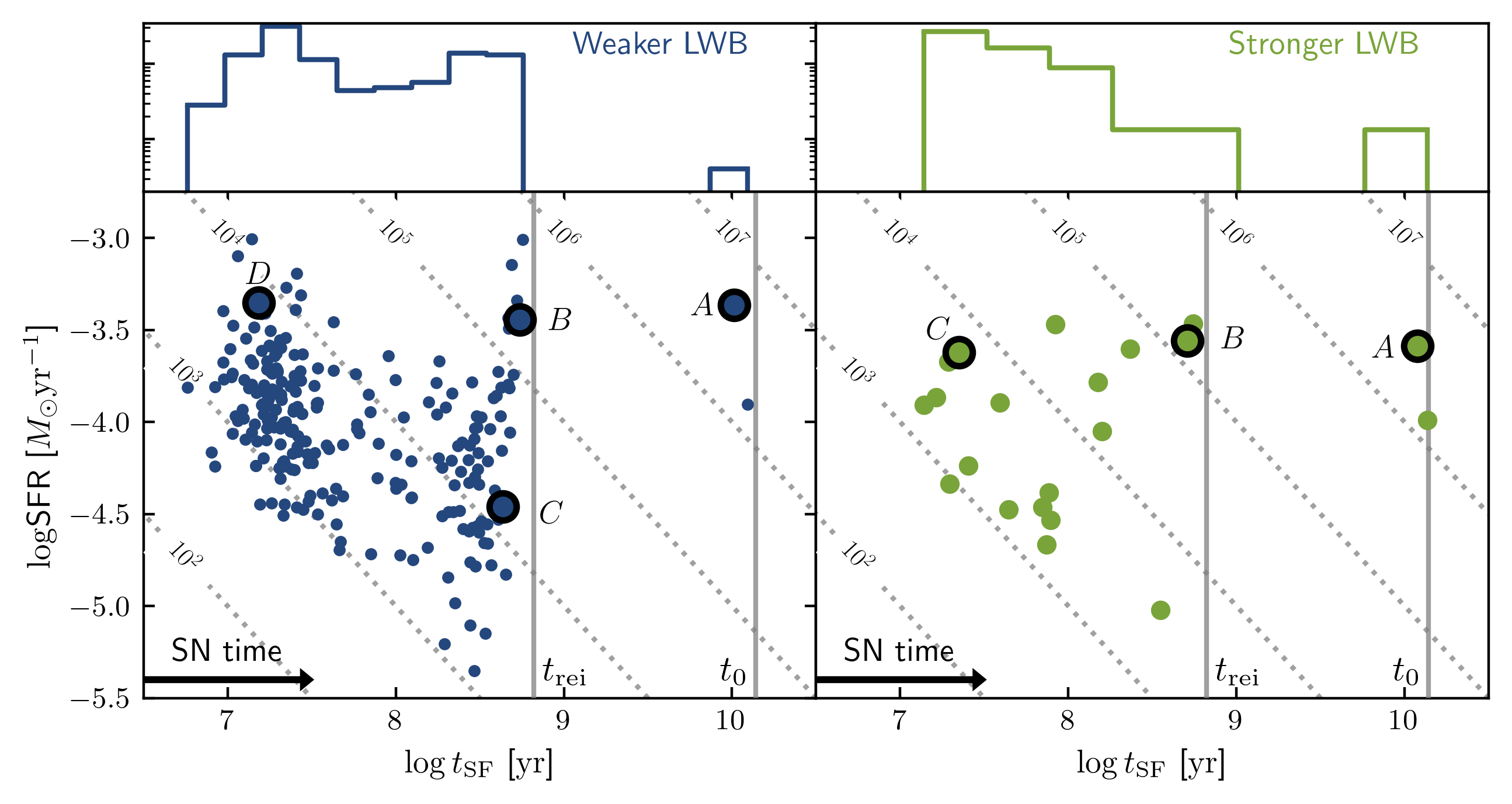}
    \caption{Distribution of the time duration over which galaxies are star forming, calculated as the difference between the oldest and youngest star, $t_{\rm{SF}}$, and the mean star formation rate during this time. Contours of constant stellar mass formed are shown for reference as the dashed diagonal lines. The two vertical grey lines show the age of the Universe at reionisation and today. Each scatter point represents an individual galaxy. The diagonal line shows contours of constant stellar mass, in \Msun. The two LWBs exhibit similar features. Both have significant populations quenching on a supernova timescale, $t_{\rm{SF}} \sim 10^{7}\rm{yr}$, at the epoch of reionisation, $t_{\rm{SF}} \sim 10^{8.5}\rm{yr}$, and those that continue star forming for the age of the universe, $t_{\rm{SF}} \sim 10^{10}\rm{yr}$. The weaker LWB shows a clear bimodal distribution between the supernova and reionisation timescale, while the stronger LWB exhibits a more continuous distribution. The minimum star formation timescale, along with a typical SFR, is the origin of the floor in the stellar mass-halo mass relation (Fig.~\ref{fig:SMHM}).}
    \label{fig:SFR_time}
\end{figure*}

A useful way of characterising the SF histories for the whole sample of galaxies is to study the duration of time which they are star forming and their average star formation rate (SFR); these are shown in Fig.~\ref{fig:SFR_time}. We calculate the time the galaxy is star forming from the difference between the oldest and youngest in situ stars, $t_{\rm{SF}}$, and the mean SFR as the birth mass of in situ stars divided by this time. These two numbers are not enough to fully describe the observed SF histories (e.g. Fig.~\ref{fig:SF_hist_individual}), in particular ignoring long quiescent periods between distinct episodes of SF, but they are still a useful way of summarising the stellar growth. The histograms in the top panels of Fig.~\ref{fig:SFR_time} show the number counts on a log scale, with an arbitrary normalisation, for $t_{\rm{SF}}$. We also plot a number of reference lines: the diagonal lines show constant stellar mass formed, the age of the universe at reionisation and today are shown with the vertical lines, and the range of stellar ages for SN progenitors is shown as the black horizontal arrow ($10^{6.5}$--$10^{7.5}\rm{yrs}$).

Considering the overall sample of $t_{\rm{SF}}$, we observe three broadly distinct populations. One quenching on timescales comparable to the lifetimes of SN progenitors, $\sim 10^{7}\rm{yr}$, one on a timescale comparable to the age of the Universe at reionisation, $\sim 10^{8.5}\rm{yr}$, and one close to a Hubble time $\sim 10^{10} \rm{yr}$. The first two populations arise from two distinct quenching mechanisms: self-quenching due to SN feedback and quenching by the external ionising background. The third population represents those galaxies that rejuvenate post-reionisation or were never quenched by reionisation.

In the weaker LWB (left panel), these three populations are very distinct, with the distribution of $t_{\rm{SF}}$ showing clear peaks at $\approx 10^{7.3}\, \rm{yr}$ and $\approx 10^{8.7}\, \rm{yr}$. We observe a large diversity of mean SFRs. For the shorter star forming population, that self-quenched on the SN timescale, the SFRs are centred on $\rm{SFR} \sim 10^{-4}$\Msun$\rm{yr}^{-1}$, with an approximate scatter of half a dex. For the longer star forming population, which are quenched by reionisation, we see a much larger distribution with $\rm{SFR} \sim 10^{-5.5}$--$10^{-3}$\Msun$\rm{yr}^{-1}$. These galaxies with low mean SFRs form through multiple epochs, but not sustained, star formation (e.g. Halo C in the weaker LWB, Fig.~\ref{fig:SF_hist_individual}). The higher SFRs correspond to haloes that are able to nearly continually form stars up to reionisation (e.g. Halo B in the weaker LWB, Fig.~\ref{fig:SF_hist_individual}).

In the stronger LWB (right panel) there is no clear distinction between a population quenching on a supernova timescale and one at reionisation. More massive galaxies ($M_{\ast} \gtrsim10^5$\Msun) are broadly unaffected by the LWB, while the fainter galaxies no longer exhibit a bimodality in $t_{\rm{SF}}$. While there are still galaxies being quenched on similar timescales to the weaker LW model there are more galaxies at intermediate SF timescales longer than the lifetime of massive stars, but notably shorter than reionisation. This is due to the introduction of an additional quenching mechanism, where galaxies can be quenched by the time-dependent LWB.

We also highlight where the four case-study haloes lie in this space. Here, we see that variations to the LWB can move galaxies from one population to another (e.g. Halo C). Other haloes only moderately change their location in this parameter space (e.g. Halo B).

\subsubsection{Analytic arguments} \label{sec:analytic_arg}

The behaviour of the galaxies forming from a single episode of star formation can be understood from some relatively simple analytic arguments.

We begin by considering the potential energy of the host halo, assuming an NFW profile \citep{NFW}. The density profile can be written as
\begin{equation}
    \rho (r) = \frac{\rho_0}{r/r_{s}(1+r/r_s)^2}.
\end{equation}
With the total potential depth of an NFW halo being
\begin{equation}
    \phi_{\rm{Bind}} = G (4/3 \pi \cdot 200 \rho_{\rm{crit}})^{1/3} M_{\rm{200c}}^{2/3} f(c),
\end{equation}
where $\rho_{\rm{crit}}$ is the critical density of the universe, and $c$ the halo concentration $c = r_{s}/R_{\rm{200c}}$ with
\begin{equation}
    f(c) = \frac{1}{c} \bigg (\ln(1+c) -\frac{c}{c+1} \bigg ).
\end{equation}
The energy required to liberate all gas from a NFW halo can be approximated as
\begin{equation}
E_{\rm{bind}} = f_b M_{\rm{200c}} \phi_{\rm{bind}},
\end{equation}
assuming haloes contain the gas mass expected from the cosmic baryon fraction, $f_b$, initially distributed at the halo centre. As supernovae are the main source of energy in this regime, it is useful to compare the binding energy of the gas to the typical energy of a supernova. The number of SN required to fully evacuate all gas from an NFW potential can then be approximated as follows
\begin{equation}
\label{Eqn:binding_energy}
\begin{aligned}
    N_{\rm{SN}} & = \frac{\epsilon E_{\mathrm{SN}}}{E_{\rm{Bind}}} \\
       		 & = 0.04 \bigg(\frac{\epsilon}{0.1} \bigg )
       		 \bigg (\frac{E_{\rm{SN}}}{10^{51}\rm{erg}} \bigg )
       		 \bigg ( \frac{f_{\rm{b}}}{0.15} \bigg ) ^{-1}
       		 \bigg ( \frac{M_{\rm{200c}}}{10^{5.5} \rm{M_{\odot}}} \bigg )^{-5/3}
       		 \bigg ( \frac{f(c)}{f(5)}\bigg ),
\end{aligned}
\end{equation}
where $E_{\mathrm{SN}}$ is the expected energy from one supernova, and $\epsilon$ the coupling efficiency. By substituting typical values for these constants ($f_{\rm{b}} = 0.15$, $E_{\mathrm{SN}} = 10^{51} \,\rm{erg}$, $\epsilon = 0.1$)\footnote{Analytic calculations suggest $\approx 28 \%$ of a SN energy is converted into the kinetic energy of the outflow \citep{Kim_15}.} and the halo mass at which haloes first form their stars (\Mhalo$=10^{5.5}$\Msun, see Fig.~\ref{fig:formation_mass}) we find the number of supernovae required to evacuate a halo of gas to be $N_{\mathrm{SN}}= 0.04$.\footnote{We additionally assume $c=5$, which is typical of the haloes in our sample at $z=20$, and consistent with models for the concentration mass relation \citep[e.g.][]{Correa_15,Ludlow_16,Diemer_19,Brown_22}. We also note that the dependence on halo concentration is relatively mild, only changing the inferred numbers by order unity.} Therefore, most of the haloes in our sample begin forming stars while they are far too small to self-regulate, with the first supernova able to completely shut down star formation and remove all gas. The above calculation suggests that a halo needs to be at least as massive as \Mhalo{} $\gtrsim 10^{6.5}$\Msun{} for it to retain some gas after multiple supernovae.

\begin{figure}
    \centering
    \includegraphics[width=\linewidth]{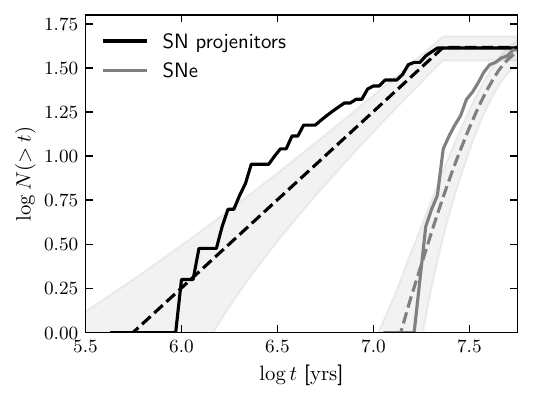}
    \caption{The cumulative number of SNe since star formation began for Halo C in the stronger LWB. This galaxy forms from a single burst of stars. The solid black line shows when SN progenitors are born, and the solid grey line when SN occur. The analytic expectation for a constant star formation rate is shown as dashed lines, with Poisson noise (see Eqn.~(\ref{Eqn:SN_num1}--\ref{Eqn:SN_num2})), and matches the simulation well. The first massive star that will die as a SN is not born for $\sim 1 \rm{Myr}$, and the first SN occurs at $\sim 10\rm{Myr}$, directly quenching the system.}
    \label{fig:SN_time}
\end{figure}

As discussed in Section~\ref{Sec:Methods}, one of the novel aspects of these simulations is the modelling of individual supernovae from a stochastically sampled IMF. We are therefore able to track the exact number of supernovae and the time they occur within the simulations. In Fig.~\ref{fig:SN_time} we show when the SN progenitors are born (black solid line) and subsequently die (grey solid line) for Halo C with the stronger LWB, which is typical of galaxies forming from an single episode of star formation. Here, halo C forms $M_{\ast} = 5 \times 10^3$\Msun{} and is star forming for $t_{\rm{sf}} = 2 \times 10^{7}\rm{yr}$, with a mean $\rm{SFR} = 2 \times 10^{-4}$\Msun$\rm{yr}^{-1}$. It is able to form $51$ massive stars ($>8$\Msun), $41$ of which end their life as supernovae (the other massive stars collapse directly to stellar mass black holes without ejecting energy). From Fig.~\ref{fig:SN_time}, it is clear that star formation ceases as soon as the first SN occurs, with subsequent SNe happening for another $\approx 30\, \rm{Myr}$. We observe that for this galaxy it takes $\sim 10^6 \rm{yr}$ for the first SN progenitors to be born, which are statistically more likely to be less massive ($\sim 8$\Msun) with lifetimes $\sim 10^{7.5}\rm{yr}$, with massive stars continuing to form until the galaxy becomes quenched. The first supernova occurs at $t \sim 10^{7}\rm{yr}$, after which time the system ceases to form stars.

We recall that the stellar lifetimes of massive stars are given by \cite{Portinari_98}, which range from $10^{6.5}$--$10^{7.5}\rm{yr}$, and we assume a \cite{Kroupa_01} IMF. The energy and mass returns are taken from \cite{Sukhbold_16}, where the initial stellar mass of the star dictates if it dies as a SN or a direct collapse black hole or neutron star. While this is solely dictated by the stellar birth mass, the relation is highly non-trivial with disconnected mass ranges being predicted to become SN (see for example their Fig.~8).

The production of SN progenitors is related to the SFR by
\begin{equation}\label{Eqn:SN_num1}
    \frac{d n_{\rm{Born}}}{dt} = \int^{m_{\rm{max}}}_{m_{\rm{min}}} \rm{SFR}(t) \xi(m)f(m) \ dm
\end{equation}
and the number of SN given by
\begin{equation}
    \frac{d n_{\rm{SN}}}{dt} = \int^{m_{\rm{max}}}_{m_{\rm{min}}} \rm{SFR}(t-t_{\rm{age}}(m)) \xi(m)f(m) \ dm.
\end{equation}
Here $\xi$ is the IMF, $m_{\rm{min}}$ and $m_{\rm{max}}$ the smallest and largest stars that will die as SN, which we take $m_{\rm{min}} = 8$\Msun$\,$ and $m_{\rm{max}} = 100$\Msun. $f(m)$ is a binary function, taking on a value of $1$ if the star will die as a SN, and $0$ if it will become a direct collapse stellar mass black hole or neutron star. $t_{\rm{age}}$ is the lifetime of different stars, and is given by the star's initial mass.\footnote{The stellar ages from \cite{Portinari_98} exhibit a mild metallicity dependence. For simplicity, we take the stellar ages for $\log Z/Z_{\odot} = -4$.}

In Fig~\ref{fig:SN_time} we show these analytic calculations as dashed lines assuming a top-hat SF history of the form
\begin{equation}\label{Eqn:SN_num2}
    \rm{SFR}(t) =
    \begin{cases}
   	 \mathrm{SFR_0},& \text{if }\ 0<t<t_{\rm{SF}} \\
   	 0,& \text{otherwise}
    \end{cases}.
\end{equation}
To match the galaxy being considered, we use $t_{\rm{SF}} = 2 \times 10^{7}\rm{yr}$ and $\mathrm{SFR_0} = 2 \times 10^{-4}$\Msun$\rm{yr}^{-1}$. The grey bands represent the Poisson counting error. These calculations closely follow what is observed in the simulations, providing the correct prediction for when the first massive stars are born and die as SN, as well as the correct number of total SN. Differences between the simulations and the analytic calculations are a combination of the Poisson noise, and the fact that the SF histories are not exactly  constant.

From the above calculations, we see a natural explanation for the very specific star forming timescale observed in Fig~\ref{fig:SFR_time}. For the typically observed SFRs ($\sim 10^{-3}$--$10^{-4}$\Msun$\rm{yr}^{-1}$), it is the typical time between star formation beginning and the first SN occurring ($t \sim 10^7\rm{yr}$), which can be derived from the above equations. We note that in the current LYRA model we do not implement pre-supernovae feedback, and as such do not have feedback, or energy injection, prior to a star becoming a supernovae. While we do not expect this to effect the qualitative picture outlined above, in particular that the very first supernovae are able to fully evacuate the gas from these small haloes, it will likely change the quantitative prediction for the stellar mass, and potentially the exact quenching timescales for these systems.

\subsection{The origin of the SMHM relation}

Here we summarise how the results from Section~\ref{sec:critical_mass} \& \ref{Sec:SF_histories} lead to the the distinct features observed in the stellar mass halo mass relation for our simulations (Section~\ref{sec:SMHM}, Fig.~\ref{fig:SMHM}). Firstly, all galaxies in our sample form their first stars at similar halo masses (Fig.~\ref{fig:formation_mass}), leading to comparable star formation rates during these first episodes of star formation (e.g. Fig~\ref{fig:SF_hist_individual}). Due to the small halo masses when our galaxies first form stars (\Mhalo$\sim 10^{5.5}$\Msun, Fig.~\ref{fig:formation_mass}) they are unable to self-regulate, and are quenched by their first SN (Eqn.~\ref{Eqn:binding_energy}, Fig.~\ref{fig:SN_time}). This leads to a characteristic minimum time scale for star formation of $t_{\rm{SF}} \sim 10^{7} \rm{yr}$, dictated by the assumed stellar ages and IMF as well as the star formation rate. For small haloes, this is the only star formation during their lifetimes and sets the clear floor to the SMHM relation at $M_{\rm{st}} \sim 10^3$--$10^4$\Msun.

In the weaker LWB there is a clear multimodal distribution of star formation times, corresponding to being quenched by either SN or reionisation. However, this does not manifest as a clear break, or multimodal populations, in the SMHM. Instead, these galaxies also grow significantly through accretions, leading to a smooth SMHM relation that is extremely shallow, almost constant, at low halo masses before steepening for large haloes. For the stronger LW model, a number of galaxies are also quenched by the time varying LWB. This leads to a population of haloes with constant stellar mass at low masses (\Mhalo{} $\lesssim 10^9$\Msun), forming from a single burst of stars. We see a clear break in SMHM at \Mhalo$\sim10^9$\Msun, with both these small galaxies ($M_{\ast} \sim 10^3$\Msun) and others that were massive enough to overcome the LWB at high redshift and continue forming stars up to reionisation ($M_{\ast} \sim 10^5$\Msun).

\section{Discussion} \label{Sec:discussion}

\subsection{Potential Caveats}

While the presented model has many strengths, in particular the very high resolution ($4$\Msun) and treatment of individual SN, there are some physical processes that are missing within the simulations. Here, we discuss what we believe to be the most important ones for the dwarfs in our sample, and which we hope to explore in future work.

In this work we have assumed a spatially uniform, but temporally evolving, LWB. It is expected for there to be an environmental dependence to this background. From \cite{Inc23} the distribution of LW intensities is highly skewed to higher intensities, with most of the Universe being close to the mean value but with high density environments being subject to almost two orders of magnitude higher intensity. This complicates the implementation of the `correct' LWB, where one should attempt to reproduce the LWB to best match their given observations. For example, when comparing to Milky Way satellites these will be subject to the LWB in the region of the proto-Milky Way.

We also have not attempted to produce a self consistent LWB. Clearly, in the somewhat extreme case of the weaker LWB where there is essentially no LWB before reionisation (see Fig.~\ref{fig:LW_background}) the results from our own simulations would predict a stronger LWB than the one we assume. For the stronger LWB it is unclear if we have this inconsistently at certain redshifts. While modelling the contribution to the LWB from a population of dwarf galaxies is beyond the scope of this paper, this is a potentially useful way of placing some self-consistent constraints on the very early LWB in future models.

When generating the initial conditions for our simulations we have assumed that DM and baryons share the same initial density and velocity distribution, as is common for most cosmological simulations. This approach ensures that the total matter power spectrum is consistent with the assumed cosmology, but it is not correct in detail. It is instead expected that the baryonic density field will be systematically underdense when compared to DM (known as baryon-DM isocurvature perturbations). Large scale coherent differences between these two density fields imprint a coherent velocity on large scales ($\sim$Mpc) known as baryonic streaming velocities \citep{Tseliakhovich}. The effects of baryon-DM isocurvature perturbations have yet to be studied in the dwarf galaxy regime. Only in recent years have the methods been developed to accurately imprint these small differences in the initial conditions \citep[e.g.][]{Hahn_21}, with recent work using the Flamingo simulations show that this effect can suppress star formation by $\sim 20$\% at $z=8$, with the larger effects at earlier times (Jessop et al. \textit{in prep}). The streaming velocities have been studied much more within the literature, and work to suppress and delay star formation in haloes. It is expected that this effect will be most important at very high redshift ($z \gtrsim 15$) \citep{Schauer_19, Schauer_21}, and will act to delay the formation of the very first stars by $\sim 50\,\rm{Myr}$. However, recent works suggest that the effect on dwarf galaxies is small, if not negligible. \cite{Schauer_23} find no measurable effect of the dwarf galaxy populations and their halo occupation fraction by $z=5$. \cite{Nadler_25} suggest there may be a modest impact on the present day occupation fraction, moving the transition from dark to luminous by $\sim 0.2\, \rm{dex}$.

In our star formation model we use a density and temperature threshold, as is common for many similar simulations. Specifically requiring gas to be denser then $n_{\mathrm{H}} > 10^3 \mathrm{cm^{-3}}$ and colder than $T < 100\mathrm{K}$. We can then wonder if these specific choices significantly impact our results. From our study of the gas phase space during the first epochs of star formation (Fig.~\ref{fig:first_stars_halo_134} \& \ref{fig:Gas_first_stars_halo018}) we can see that the gas is able to cool below $T = 100\mathrm{K}$ before crossing the $n_{\mathrm{H}} = 10^3 \mathrm{cm^{-3}}$, and as such star formation at this time is dictated by our choice of density threshold. From the shape of the cooling function it is clear that once the gas reaches this density it will, in the absence of feedback, eventually form a star. As the free fall time of this gas is short (e.g. for $n_{\mathrm{H}} = 10^3 \mathrm{cm^{-3}}$, $t_{\mathrm{ff}} \approx 5 \times 10^5\mathrm{yr}$, while for $n_{\mathrm{H}} = 10 \mathrm{cm^{-3}}$ $t_{\mathrm{ff}} \approx 10^7\mathrm{yr}$), changing the density threshold will not meaningfully change star formation on a cosmic timescale. As such, we conclude that our star formation criterion will not meaningfully change which haloes are able to form stars or the halo mass and redshift at which they do so. Similarly, we do not expect feedback to be significantly impacted by this choice, especially for the small haloes that self quench from their first SNe.

Currently, we do not implement a specific PopIII star formation model. While we do form stars at primordial metallicities, we do not treat their formation and evolution differently, importantly assuming a standard \cite{Kroupa_01} IMF when sampling stellar masses. The true properties and evolution of these metal-free stars is highly uncertain \citep[see][for a review]{PopIII_review}. While it is quite likely they form from a non-standard IMF, it is unclear what form it takes and the minimum and maximum mass of these stars. Even more uncertain is how these stars end their lives, with some models suggesting that they all become direct collapse black holes or alternately as particularly energetic supernovae. It is unclear to what extent the properties of present day dwarfs are sensitive to these choices. However, some of our systems, such as those forming from a single burst of star formation at high redshift, are likely sensitive to these different assumptions and are promising probes to explore the properties of these first stars. We hope to explore a range of PopIII models in future work, making predictions for their effect on the properties of our dwarf galaxies today.

In the current version of the LYRA model we do not have on-the-fly radiative transfer, and do not implement an effective radiative feedback model prior to a star becoming a SN. We instead inject all energy from the energy yields \citep{Sukhbold_16} at the end of a star's life, ignoring pre-supernova processes. This likely means that the star formation histories of our dwarfs are overly bursty \citep[see for example][]{Edge_2}. At the low mass end, it may also mean that we are forming too many stars, and potentially too few in more massive haloes. In addition, we currently only use non-equilibrium cooling rates for $\rm{HI}$, $\rm{HII}$, $\rm{HeI}$, $\rm{HeII}$ and $\rm{HeIII}$, with equilibrium rates for metals and $\mathrm{H_{2}}$. The complex coupling between these two processes and its subsequent effect on feedback in regulating galaxy formation remains unclear.

\subsection{The need for a cold ISM to model faint dwarf galaxies}

We have shown that star formation may begin in much smaller haloes than assumed in other works, and depends strongly on the assumed LWB. This is most clearly shown in Fig.~\ref{fig:formation_mass} where almost all galaxies in our sample, for both LWBs, form their first stars when \Mhalo$\sim 10^{5}$--$10^{6}$\Msun, far below the atomic cooling limit. In Fig.~\ref{fig:first_stars_halo_134}, we show the gas begins collapsing from moderate densities, $n_{\rm{H}} \sim 0.01$--$1\, \rm{cm}^{-3}$, and relatively cool temperatures, $T \sim 100$--$1,000 \, \rm{K}$.

It is clear that successfully modelling low mass dwarf galaxies ($M_{\rm{st}} \lesssim 10^{6}$\Msun) requires both high mass resolution and directly modelling the cold gas phase, including the contribution of $\rm{H_{2}}$ cooling. In addition, stochastically sampling the IMF is important to ensure the correct timescales for quenching (see Fig~\ref{fig:SN_time}) and the onset of stellar feedback. Many cosmological simulations, both zooms and boxes, have only a moderate resolution, and utilise an effective equation of state. Such simulations will artificially force their dwarf galaxies to form stars at later times and in larger haloes than they physically should. See \cite{Apostle_sf_thresh} for an example in the EAGLE model, where they show how the mass threshold for star formation before reionisation is set by the assumed equation of state and star formation threshold.

\subsection{RELHICs and primordial metallicity Lyman limit systems}

It is expected that haloes of mass $10^8 \lesssim$\Mhalo$/$\Msun$\lesssim 10^9$ today are able to overcome the background UV field and form a reservoir of neutral gas, but remain too small to efficiently cool this gas and begin star formation. Interestingly, models that only consider atomic $\rm{H}$ cooling predict that the present day occupation fraction transitions from fully dark to luminous at \Mhalo$\sim 10^9$\Msun (e.g. Fig.~\ref{fig:Occupation_fraction}). This leads to the prediction of DM haloes with clouds of neutral hydrogen that do not host a galaxy. These REionisation-Limited HI Clouds (RELHICs) \citep[e.g.][]{Llambay_17}, or similarly Lyman limit systems with primordial metallicity, would be observable through their HI 21cm emission or the HI absorption in a background source's spectrum.

As shown in Fig.~\ref{fig:Occupation_fraction}, if the cold gas is directly modelled, and $\rm{H_{2}}$ cooling considered, the transition from dark to luminous haloes occurs at notably smaller masses: \Mhalo{} $\sim 10^7$--$10^8$\Msun, depending on the strength of the LWB, with inferences from Milky Way satellites suggesting it is \Mhalo{} $\sim 10^8$\Msun. This is below the mass in which we expect haloes to host neutral gas today. Therefore, starless, pristine, HI clouds may not exist in the local Universe, or may be exceedingly rare so as to have none that are observable today. Instead, many low mass, metal-poor HI clouds could likely host a dwarf galaxy at their centre, though the light from such small systems may not currently be detectable with reasonable exposure times.

A recent RELHIC candidate is that of Cloud-9 \citep{Zhou_23} - a gas cloud of neutral gas detected in 21cm emission near the galaxy M94 with no stellar component currently detected. Follow-up VLA-D observations were presented in \cite{Llambay_24}, concluding that, while the RELHIC hypothesis can not currently be ruled out, it seems most likely that the systems hosts a faint galaxy ($M_{\ast}\lesssim 10^5$) that is below the surface brightness limit of current surveys, such as DESI. Indeed, with the inferred halo mass of this system (\Mhalo$\sim 10^{9.5}$--$10^{10}$\Msun{}) our own results would suggest this system should not be dark.

The existence of such systems is currently an open question and searches for these metal-poor $\rm{HI}$ systems will continue. Their detection, or lack thereof, would provide powerful constraints on galaxy formation in the early Universe.

\section{Conclusion} \label{Sec:summary}

We have presented a new suite of dwarf galaxies simulated with the LYRA galaxy formation model using an extremely high resolution of $4$\Msun{}. The new suite consists of $65$ systems in the halo mass range \Mhalo$\approx10^{7}$--$5 \times 10^{9}$\Msun with stellar masses $M_{\ast} \approx 100$--$5\times 10^6$\Msun. Haloes are chosen from Local Group-like environments, and sample a diverse range of accretion histories and analogues to field dwarfs observable in our own Universe. The new sample probes the transition from classical dwarfs to ultra faints and small haloes unable to form stars.

In this work we consider the effect of two different background radiation fields in the early Universe ($z > 7$), differing in the assumed Lyman-Werner (LW) intensity. LW photons have energies $E\sim10$--$13\rm{eV}$; while non-ionising, they are able to efficiently photodissociate and destroy molecular hydrogen, a key cooling channel for metal-poor gas. The strength of this field in the early Universe is highly uncertain (see Fig.~\ref{fig:LW_background}). Our fiducial model assumes the LWB from the tabulated \cite{FG20} spectra, where the strength of the LW is very weak in the early Universe before significantly increasing at reionisation. Theoretical models of the LWB suggest this behaviour is unlikely. We therefore also model the effects of using a more slowly varying LWB, taking the functional form from \cite{Inc23}. The two background spectra only differ for $z> 7$, and share the same intensity for the ionising part of the spectrum (see Section~\ref{section:cooling_tables} and Fig.~\ref{fig:LW_background}). Our main results can be summarised as follows.

(i) We present the stellar mass-halo mass (SMHM) relation in Fig.~\ref{fig:SMHM}. We find classical dwarfs, $M_{\ast} > 10^5$\Msun, are only moderately impacted by the different LWBs and are hosted by haloes with \Mhalo{} $\gtrsim 10^9$\Msun{}. Smaller ultra-faint dwarf galaxies, $M_{\ast} < 10^5$\Msun, are strongly affected by the different LWBs. In both cases there is a clear floor in stellar mass, centred at $M_{\ast} \sim 10^3$\Msun{}, with notable scatter ($\approx 0.2\, \rm{dex}$). The weaker LWB has a smooth and relatively shallow SMHM relation at low masses, with ultra-faint dwarfs being hosted by haloes with a wide range of masses. On the other hand, the stronger LWB exhibits a sharp transition in the SMHM relation at \Mhalo$\sim 10^9$\Msun, with galaxies of $M_{\ast} \sim 10^3$ and $\sim 10^5$\Msun{} hosted in comparable mass haloes.

(ii) The mass scale at which haloes transition from being dark, and unable to form stars, to hosting a galaxy is dramatically affected by strength of the early LW radiation. This is quantified through the halo occupation fraction (Fig.~\ref{fig:Occupation_fraction}). In the weaker LWB the transition occurs at \Mhalo{} $\approx 10^{7}$\Msun, while for the stronger LWB this is increased to \Mhalo{} $\approx 10^{8}$\Msun. Both LWBs predict the transition to occur at significantly smaller masses than models that only consider atomic cooling, highlighting the importance of $\rm{H_2}$ cooling when modelling faint dwarf galaxies. Our stronger LWB is roughly consistent (within a factor of a few) with models that fit the Milky Way's satellite population, while the transition in the weak LW is at significantly smaller mass than all other models from the literature.

(iii) We explore at what time and halo mass our galaxies first form stars in Fig.~\ref{fig:formation_mass}. All systems in our sample first form stars prior to reionisation, and as early as $z\approx 25$. In the weaker LWB, haloes initiate star formation when they cross a mass threshold of \Mhalo{} $\approx 10^5$--$10^6$\Msun, with haloes continually beginning star formation up to cosmic reionisation ($z\sim 7)$. For the stronger LWB, haloes begin forming stars at a similar mass, but this is  restricted to $z \gtrsim 15$, with the stronger LWB preventing the smallest haloes from forming stars.

(iv) We study the correlation between the star formation efficiency of a halo and its formation time in Fig.~\ref{fig:Bias_accretion}. In the weaker LWB we find a strong correlation with formation time; dark haloes form comparably late and are undermassive at high redshift compared to their luminous counterparts, with luminous haloes that are overmassive at reionisation are able to form more stars. Interestingly, in the stronger LWB this correlation with formation time is not obviously present in our sample.

(v) We observe a diverse range of star formation histories (Fig.~\ref{fig:SF_hist_individual}). Reionisation `survivors' continue through and after the epoch of reionisation (e.g. Halo A). Reionisation relics continue to form stars up to reionisation, but are permanently quenched afterwards (e.g. Halo B). Some systems are quenched prior to reionisation: some self-quench due to SN feedback and form from a single burst of stars (Halo D, weaker LW), while others are quenched by the LWB (halo C, stronger LW).

(vi) Many galaxies form from a single burst of stars prior to reionisation, with a minimum star formation time of $\sim 10^7\, \rm{yr}$ (Fig.~\ref{fig:SFR_time}). This population leads to the observed floor in the SMHM relation at $M_{\ast} \sim 10^3$\Msun. This characteristic minimum time scale corresponds to the time from star formation igniting to the first SN occurring, which are able to completely evacuate haloes of their gas (see Fig.~\ref{fig:SN_time} and equations within Section~\ref{sec:analytic_arg}). We also observe distinct populations of galaxies that are quenched by cosmic reionisation, which is most easily observed in the weaker LWB.

We have shown that the properties, and expected number, of ultra faint dwarf galaxies ($M_{\ast} < 10^5$\Msun) are very sensitive to the assumed LWB at high redshift. The main source of this background comes from stars, with PopIII star formation expected to make up a significant component at early times. While it is already well established that dwarf galaxies are significantly impacted by cosmic reionisation, we have shown that the ultra faint population is also sensitive to the assumed properties of the Universe far before this epoch. The faintest galaxies therefore offer a window into the Universe before reionisation, and are promising probes to constrain the  Universe's early ($z \sim 10$) star formation rate and the properties of the first stars.  

There will soon be a wealth of new data in the low surface brightness regime, with surveys such as DESI, Euclid and LSST expected to find and characterise countless more dwarf galaxies in our local Universe. High resolution simulations that sample a wide range of halo masses, accretion histories and variations in uncertain physical parameters, such as those presented in this work, will be key to interpreting these upcoming observations. With a concerted effort from both the observational and theoretical communities, we are poised to make great progress in dwarf galaxy science in the coming decade, and can use these systems as probes to further study the properties of the very early Universe, constrain our cosmological model, and further understand the role of feedback in driving galaxy formation.

\section*{Acknowledgments}

The authors thank Kyle Oman, Alexander Riley, Isabel Santos-Santos and Sadegh Khochfar for useful discussion throughout the project, Robert Grand for access to the Auriga simulation data and Alex Richings for help generating the necessary CHIMES files for the cooling tables. S.T.B. and A.F. acknowledge support by a UK Research and Innovation (UKRI) Future Leaders Fellowship (grant no MR/T042362/1) and a Sweden's Wallenberg Academy Fellowship. S.P. acknowledges funding by the Austrian Science Fund (FWF) through grant-DOI: 10.55776/V982. S.B. is supported by the UKRI Future Leaders Fellowship (grant numbers MR/V023381/1 and UKRI2044). J.S. acknowledges support from the Science and Technologies Facilities Council (STFC) through the studentship grant ST/X508354/1. J.E.D. is supported by the United Kingdom Research and Innovation (UKRI) Future Leaders Fellowship `Using Cosmic Beasts to uncover the Nature of Dark Matter' (grant number MR/X006069/1). A.R.J. is supported by the STFC consolidated grant ST/X001075/1. This work used the DiRAC@Durham facility managed by the Institute for Computational Cosmology on behalf of the STFC DiRAC HPC Facility (www.dirac.ac.uk). The equipment was funded by BEIS capital funding via STFC capital grants ST/K00042X/1, ST/P002293/1, ST/R002371/1 and ST/S002502/1, Durham University and STFC operations grant ST/R000832/1. DiRAC is part of the National e-Infrastructure.

%%%%%%%%%%%%%%%%%%%%%%%%%%%%%%%%%%%%%%%%%%%%%%%%%%
\section*{Data Availability}

The simulations and data presented here are available upon a reasonable request to the corresponding author.



\bibliographystyle{mnras}
\bibliography{example} % if your bibtex file is called example.bib

\begin{thebibliography}{}
\makeatletter
\relax
\def\mn@urlcharsother{\let\do\@makeother \do\$\do\&\do\#\do\^\do\_\do\%\do\~}
\def\mn@doi{\begingroup\mn@urlcharsother \@ifnextchar [ {\mn@doi@} {\mn@doi@[]}}
\def\mn@doi@[#1]#2{\def\@tempa{#1}\ifx\@tempa\@empty \href {http://dx.doi.org/#2} {doi:#2}\else \href {http://dx.doi.org/#2} {#1}\fi \endgroup}
\def\mn@eprint#1#2{\mn@eprint@#1:#2::\@nil}
\def\mn@eprint@arXiv#1{\href {http://arxiv.org/abs/#1} {{\tt arXiv:#1}}}
\def\mn@eprint@dblp#1{\href {http://dblp.uni-trier.de/rec/bibtex/#1.xml} {dblp:#1}}
\def\mn@eprint@#1:#2:#3:#4\@nil{\def\@tempa {#1}\def\@tempb {#2}\def\@tempc {#3}\ifx \@tempc \@empty \let \@tempc \@tempb \let \@tempb \@tempa \fi \ifx \@tempb \@empty \def\@tempb {arXiv}\fi \@ifundefined {mn@eprint@\@tempb}{\@tempb:\@tempc}{\expandafter \expandafter \csname mn@eprint@\@tempb\endcsname \expandafter{\@tempc}}}

\bibitem[\protect\citeauthoryear{{Abel}, {Anninos}, {Zhang}  \& {Norman}}{{Abel} et~al.}{1997}]{Abel_97}
{Abel} T.,  {Anninos} P.,  {Zhang} Y.,   {Norman} M.~L.,  1997, \mn@doi [\na] {10.1016/S1384-1076(97)00010-9}, \href {https://ui.adsabs.harvard.edu/abs/1997NewA....2..181A} {2, 181}

\bibitem[\protect\citeauthoryear{{Ahn}, {Shapiro}, {Iliev}, {Mellema}  \& {Pen}}{{Ahn} et~al.}{2009}]{Ahn_09}
{Ahn} K.,  {Shapiro} P.~R.,  {Iliev} I.~T.,  {Mellema} G.,   {Pen} U.-L.,  2009, \mn@doi [\apj] {10.1088/0004-637X/695/2/1430}, \href {https://ui.adsabs.harvard.edu/abs/2009ApJ...695.1430A} {695, 1430}

\bibitem[\protect\citeauthoryear{{Ahvazi}, {Benson}, {Sales}, {Nadler}, {Weerasooriya}, {Du}  \& {Bovill}}{{Ahvazi} et~al.}{2024}]{Galacticus_2}
{Ahvazi} N.,  {Benson} A.,  {Sales} L.~V.,  {Nadler} E.~O.,  {Weerasooriya} S.,  {Du} X.,   {Bovill} M.~S.,  2024, \mn@doi [\mnras] {10.1093/mnras/stae761}, \href {https://ui.adsabs.harvard.edu/abs/2024MNRAS.529.3387A} {529, 3387}

\bibitem[\protect\citeauthoryear{{Behroozi}, {Wechsler}  \& {Conroy}}{{Behroozi} et~al.}{2013}]{Behroozi_13}
{Behroozi} P.~S.,  {Wechsler} R.~H.,   {Conroy} C.,  2013, \mn@doi [\apj] {10.1088/0004-637X/770/1/57}, \href {https://ui.adsabs.harvard.edu/abs/2013ApJ...770...57B} {770, 57}

\bibitem[\protect\citeauthoryear{{Benitez-Llambay} \& {Frenk}}{{Benitez-Llambay} \& {Frenk}}{2020}]{BL_20}
{Benitez-Llambay} A.,  {Frenk} C.,  2020, \mn@doi [\mnras] {10.1093/mnras/staa2698}, \href {https://ui.adsabs.harvard.edu/abs/2020MNRAS.498.4887B} {498, 4887}

\bibitem[\protect\citeauthoryear{{Ben{\'\i}tez-Llambay} et~al.,}{{Ben{\'\i}tez-Llambay} et~al.}{2017}]{Llambay_17}
{Ben{\'\i}tez-Llambay} A.,  et~al., 2017, \mn@doi [\mnras] {10.1093/mnras/stw2982}, \href {https://ui.adsabs.harvard.edu/abs/2017MNRAS.465.3913B} {465, 3913}

\bibitem[\protect\citeauthoryear{{Ben{\'\i}tez-Llambay}, {Dutta}, {Fumagalli}  \& {Navarro}}{{Ben{\'\i}tez-Llambay} et~al.}{2024}]{Llambay_24}
{Ben{\'\i}tez-Llambay} A.,  {Dutta} R.,  {Fumagalli} M.,   {Navarro} J.~F.,  2024, \mn@doi [\apj] {10.3847/1538-4357/ad65d9}, \href {https://ui.adsabs.harvard.edu/abs/2024ApJ...973...61B} {973, 61}

\bibitem[\protect\citeauthoryear{{Benson}}{{Benson}}{2012}]{Galacticus_1}
{Benson} A.~J.,  2012, \mn@doi [\na] {10.1016/j.newast.2011.07.004}, \href {https://ui.adsabs.harvard.edu/abs/2012NewA...17..175B} {17, 175}

\bibitem[\protect\citeauthoryear{{Bertschinger}}{{Bertschinger}}{2001}]{Bertschinger_01}
{Bertschinger} E.,  2001, \mn@doi [\apjs] {10.1086/322526}, \href {https://ui.adsabs.harvard.edu/abs/2001ApJS..137....1B} {137, 1}

\bibitem[\protect\citeauthoryear{{Bose}, {Deason}  \& {Frenk}}{{Bose} et~al.}{2018}]{Bose_18}
{Bose} S.,  {Deason} A.~J.,   {Frenk} C.~S.,  2018, \mn@doi [\apj] {10.3847/1538-4357/aacbc4}, \href {https://ui.adsabs.harvard.edu/abs/2018ApJ...863..123B} {863, 123}

\bibitem[\protect\citeauthoryear{{Bose}, {Deason}, {Belokurov}  \& {Frenk}}{{Bose} et~al.}{2020}]{Bose_20}
{Bose} S.,  {Deason} A.~J.,  {Belokurov} V.,   {Frenk} C.~S.,  2020, \mn@doi [\mnras] {10.1093/mnras/staa1199}, \href {https://ui.adsabs.harvard.edu/abs/2020MNRAS.495..743B} {495, 743}

\bibitem[\protect\citeauthoryear{{Brown}, {McCarthy}, {Stafford}  \& {Font}}{{Brown} et~al.}{2022}]{Brown_22}
{Brown} S.~T.,  {McCarthy} I.~G.,  {Stafford} S.~G.,   {Font} A.~S.,  2022, \mn@doi [\mnras] {10.1093/mnras/stab3394}, \href {https://ui.adsabs.harvard.edu/abs/2022MNRAS.509.5685B} {509, 5685}

\bibitem[\protect\citeauthoryear{{Bryan} \& {Norman}}{{Bryan} \& {Norman}}{1998}]{Bryan_norman_98}
{Bryan} G.~L.,  {Norman} M.~L.,  1998, \mn@doi [\apj] {10.1086/305262}, \href {https://ui.adsabs.harvard.edu/abs/1998ApJ...495...80B} {495, 80}

\bibitem[\protect\citeauthoryear{{Chen}, {Magg}, {Hartwig}, {Glover}, {Ji}  \& {Klessen}}{{Chen} et~al.}{2022}]{A-sloth}
{Chen} L.-H.,  {Magg} M.,  {Hartwig} T.,  {Glover} S. C.~O.,  {Ji} A.~P.,   {Klessen} R.~S.,  2022, \mn@doi [\mnras] {10.1093/mnras/stac933}, \href {https://ui.adsabs.harvard.edu/abs/2022MNRAS.513..934C} {513, 934}

\bibitem[\protect\citeauthoryear{{Cole}, {Lacey}, {Baugh}  \& {Frenk}}{{Cole} et~al.}{2000}]{Galform}
{Cole} S.,  {Lacey} C.~G.,  {Baugh} C.~M.,   {Frenk} C.~S.,  2000, \mn@doi [\mnras] {10.1046/j.1365-8711.2000.03879.x}, \href {https://ui.adsabs.harvard.edu/abs/2000MNRAS.319..168C} {319, 168}

\bibitem[\protect\citeauthoryear{{Collins} \& {Read}}{{Collins} \& {Read}}{2022}]{Collins_22}
{Collins} M. L.~M.,  {Read} J.~I.,  2022, \mn@doi [Nature Astronomy] {10.1038/s41550-022-01657-4}, \href {https://ui.adsabs.harvard.edu/abs/2022NatAs...6..647C} {6, 647}

\bibitem[\protect\citeauthoryear{{Correa}, {Wyithe}, {Schaye}  \& {Duffy}}{{Correa} et~al.}{2015}]{Correa_15}
{Correa} C.~A.,  {Wyithe} J. S.~B.,  {Schaye} J.,   {Duffy} A.~R.,  2015, \mn@doi [\mnras] {10.1093/mnras/stv1363}, \href {https://ui.adsabs.harvard.edu/abs/2015MNRAS.452.1217C} {452, 1217}

\bibitem[\protect\citeauthoryear{{Crain} et~al.,}{{Crain} et~al.}{2015}]{EAGLE_2}
{Crain} R.~A.,  et~al., 2015, \mn@doi [\mnras] {10.1093/mnras/stv725}, \href {https://ui.adsabs.harvard.edu/abs/2015MNRAS.450.1937C} {450, 1937}

\bibitem[\protect\citeauthoryear{{Dav{\'e}}, {Angl{\'e}s-Alc{\'a}zar}, {Narayanan}, {Li}, {Rafieferantsoa}  \& {Appleby}}{{Dav{\'e}} et~al.}{2019}]{Simba}
{Dav{\'e}} R.,  {Angl{\'e}s-Alc{\'a}zar} D.,  {Narayanan} D.,  {Li} Q.,  {Rafieferantsoa} M.~H.,   {Appleby} S.,  2019, \mn@doi [\mnras] {10.1093/mnras/stz937}, \href {https://ui.adsabs.harvard.edu/abs/2019MNRAS.486.2827D} {486, 2827}

\bibitem[\protect\citeauthoryear{{Diemer} \& {Joyce}}{{Diemer} \& {Joyce}}{2019}]{Diemer_19}
{Diemer} B.,  {Joyce} M.,  2019, \mn@doi [\apj] {10.3847/1538-4357/aafad6}, \href {https://ui.adsabs.harvard.edu/abs/2019ApJ...871..168D} {871, 168}

\bibitem[\protect\citeauthoryear{{Diemer} \& {Kravtsov}}{{Diemer} \& {Kravtsov}}{2014}]{Splashback}
{Diemer} B.,  {Kravtsov} A.~V.,  2014, \mn@doi [\apj] {10.1088/0004-637X/789/1/1}, \href {https://ui.adsabs.harvard.edu/abs/2014ApJ...789....1D} {789, 1}

\bibitem[\protect\citeauthoryear{{Drlica-Wagner} et~al.,}{{Drlica-Wagner} et~al.}{2020a}]{Milky_census}
{Drlica-Wagner} A.,  et~al., 2020a, \mn@doi [\apj] {10.3847/1538-4357/ab7eb9}, \href {https://ui.adsabs.harvard.edu/abs/2020ApJ...893...47D} {893, 47}

\bibitem[\protect\citeauthoryear{{Drlica-Wagner} et~al.,}{{Drlica-Wagner} et~al.}{2020b}]{Sat_counts}
{Drlica-Wagner} A.,  et~al., 2020b, \mn@doi [\apj] {10.3847/1538-4357/ab7eb9}, \href {https://ui.adsabs.harvard.edu/abs/2020ApJ...893...47D} {893, 47}

\bibitem[\protect\citeauthoryear{{Fattahi} et~al.,}{{Fattahi} et~al.}{2016}]{APOSTLE_2}
{Fattahi} A.,  et~al., 2016, \mn@doi [\mnras] {10.1093/mnras/stv2970}, \href {https://ui.adsabs.harvard.edu/abs/2016MNRAS.457..844F} {457, 844}

\bibitem[\protect\citeauthoryear{{Faucher-Gigu{\`e}re}}{{Faucher-Gigu{\`e}re}}{2020}]{FG20}
{Faucher-Gigu{\`e}re} C.-A.,  2020, \mn@doi [\mnras] {10.1093/mnras/staa302}, \href {https://ui.adsabs.harvard.edu/abs/2020MNRAS.493.1614F} {493, 1614}

\bibitem[\protect\citeauthoryear{{Ferland} et~al.,}{{Ferland} et~al.}{2017}]{Cloudy}
{Ferland} G.~J.,  et~al., 2017, \mn@doi [\rmxaa] {10.48550/arXiv.1705.10877}, \href {https://ui.adsabs.harvard.edu/abs/2017RMxAA..53..385F} {53, 385}

\bibitem[\protect\citeauthoryear{{Fitts} et~al.,}{{Fitts} et~al.}{2017}]{Fitts_17}
{Fitts} A.,  et~al., 2017, \mn@doi [\mnras] {10.1093/mnras/stx1757}, \href {https://ui.adsabs.harvard.edu/abs/2017MNRAS.471.3547F} {471, 3547}

\bibitem[\protect\citeauthoryear{{Forouhar Moreno}, {Ben{\'\i}tez-Llambay}, {Cole}  \& {Frenk}}{{Forouhar Moreno} et~al.}{2022}]{Victor_22}
{Forouhar Moreno} V.~J.,  {Ben{\'\i}tez-Llambay} A.,  {Cole} S.,   {Frenk} C.,  2022, \mn@doi [\mnras] {10.1093/mnras/stac3062}, \href {https://ui.adsabs.harvard.edu/abs/2022MNRAS.517.5627F} {517, 5627}

\bibitem[\protect\citeauthoryear{{Grand} et~al.,}{{Grand} et~al.}{2021}]{Auriga_L2}
{Grand} R. J.~J.,  et~al., 2021, \mn@doi [\mnras] {10.1093/mnras/stab2492}, \href {https://ui.adsabs.harvard.edu/abs/2021MNRAS.507.4953G} {507, 4953}

\bibitem[\protect\citeauthoryear{{Gutcke}}{{Gutcke}}{2024}]{Gutcke_DM_clusters}
{Gutcke} T.~A.,  2024, \mn@doi [\apj] {10.3847/1538-4357/ad5c62}, \href {https://ui.adsabs.harvard.edu/abs/2024ApJ...971..103G} {971, 103}

\bibitem[\protect\citeauthoryear{{Gutcke}, {Pakmor}, {Naab}  \& {Springel}}{{Gutcke} et~al.}{2021}]{LYRA_1}
{Gutcke} T.~A.,  {Pakmor} R.,  {Naab} T.,   {Springel} V.,  2021, \mn@doi [\mnras] {10.1093/mnras/staa3875}, \href {https://ui.adsabs.harvard.edu/abs/2021MNRAS.501.5597G} {501, 5597}

\bibitem[\protect\citeauthoryear{{Gutcke}, {Pakmor}, {Naab}  \& {Springel}}{{Gutcke} et~al.}{2022a}]{LYRA_2}
{Gutcke} T.~A.,  {Pakmor} R.,  {Naab} T.,   {Springel} V.,  2022a, \mn@doi [\mnras] {10.1093/mnras/stac867}, \href {https://ui.adsabs.harvard.edu/abs/2022MNRAS.513.1372G} {513, 1372}

\bibitem[\protect\citeauthoryear{{Gutcke}, {Pfrommer}, {Bryan}, {Pakmor}, {Springel}  \& {Naab}}{{Gutcke} et~al.}{2022b}]{LYRA_3}
{Gutcke} T.~A.,  {Pfrommer} C.,  {Bryan} G.~L.,  {Pakmor} R.,  {Springel} V.,   {Naab} T.,  2022b, \mn@doi [\apj] {10.3847/1538-4357/aca1b4}, \href {https://ui.adsabs.harvard.edu/abs/2022ApJ...941..120G} {941, 120}

\bibitem[\protect\citeauthoryear{{Gutcke}, {Despali}, {O'Neil}, {Vogelsberger}, {Fattahi}  \& {Sanders}}{{Gutcke} et~al.}{2025}]{Gutcke2025}
{Gutcke} T.~A.,  {Despali} G.,  {O'Neil} S.,  {Vogelsberger} M.,  {Fattahi} A.,   {Sanders} D.~B.,  2025, \mn@doi [arXiv e-prints] {10.48550/arXiv.2510.05258}, \href {https://ui.adsabs.harvard.edu/abs/2025arXiv251005258G} {p. arXiv:2510.05258}

\bibitem[\protect\citeauthoryear{{Haardt} \& {Madau}}{{Haardt} \& {Madau}}{2012}]{Haardt&Madau}
{Haardt} F.,  {Madau} P.,  2012, \mn@doi [\apj] {10.1088/0004-637X/746/2/125}, \href {https://ui.adsabs.harvard.edu/abs/2012ApJ...746..125H} {746, 125}

\bibitem[\protect\citeauthoryear{{Hahn}, {Rampf}  \& {Uhlemann}}{{Hahn} et~al.}{2021}]{Hahn_21}
{Hahn} O.,  {Rampf} C.,   {Uhlemann} C.,  2021, \mn@doi [\mnras] {10.1093/mnras/staa3773}, \href {https://ui.adsabs.harvard.edu/abs/2021MNRAS.503..426H} {503, 426}

\bibitem[\protect\citeauthoryear{{Hellwing}, {Frenk}, {Cautun}, {Bose}, {Helly}, {Jenkins}, {Sawala}  \& {Cytowski}}{{Hellwing} et~al.}{2016}]{COCO}
{Hellwing} W.~A.,  {Frenk} C.~S.,  {Cautun} M.,  {Bose} S.,  {Helly} J.,  {Jenkins} A.,  {Sawala} T.,   {Cytowski} M.,  2016, \mn@doi [\mnras] {10.1093/mnras/stw214}, \href {https://ui.adsabs.harvard.edu/abs/2016MNRAS.457.3492H} {457, 3492}

\bibitem[\protect\citeauthoryear{{Incatasciato}, {Khochfar}  \& {O{\~n}orbe}}{{Incatasciato} et~al.}{2023}]{Inc23}
{Incatasciato} A.,  {Khochfar} S.,   {O{\~n}orbe} J.,  2023, \mn@doi [\mnras] {10.1093/mnras/stad1008}, \href {https://ui.adsabs.harvard.edu/abs/2023MNRAS.522..330I} {522, 330}

\bibitem[\protect\citeauthoryear{{Jenkins}}{{Jenkins}}{2010}]{Jenkins_10}
{Jenkins} A.,  2010, \mn@doi [\mnras] {10.1111/j.1365-2966.2010.16259.x}, \href {https://ui.adsabs.harvard.edu/abs/2010MNRAS.403.1859J} {403, 1859}

\bibitem[\protect\citeauthoryear{{Jenkins}}{{Jenkins}}{2013}]{Jenkins_13}
{Jenkins} A.,  2013, \mn@doi [\mnras] {10.1093/mnras/stt1154}, \href {https://ui.adsabs.harvard.edu/abs/2013MNRAS.434.2094J} {434, 2094}

\bibitem[\protect\citeauthoryear{{Jiang}, {Helly}, {Cole}  \& {Frenk}}{{Jiang} et~al.}{2014}]{Jiang_14}
{Jiang} L.,  {Helly} J.~C.,  {Cole} S.,   {Frenk} C.~S.,  2014, \mn@doi [\mnras] {10.1093/mnras/stu390}, \href {https://ui.adsabs.harvard.edu/abs/2014MNRAS.440.2115J} {440, 2115}

\bibitem[\protect\citeauthoryear{{Kaplinghat}, {Tulin}  \& {Yu}}{{Kaplinghat} et~al.}{2016}]{Kaplinghat}
{Kaplinghat} M.,  {Tulin} S.,   {Yu} H.-B.,  2016, \mn@doi [\prl] {10.1103/PhysRevLett.116.041302}, \href {https://ui.adsabs.harvard.edu/abs/2016PhRvL.116d1302K} {116, 041302}

\bibitem[\protect\citeauthoryear{{Karakas} \& {Lugaro}}{{Karakas} \& {Lugaro}}{2016}]{Karakas}
{Karakas} A.~I.,  {Lugaro} M.,  2016, \mn@doi [\apj] {10.3847/0004-637X/825/1/26}, \href {https://ui.adsabs.harvard.edu/abs/2016ApJ...825...26K} {825, 26}

\bibitem[\protect\citeauthoryear{{Kauffmann}, {White}  \& {Guiderdoni}}{{Kauffmann} et~al.}{1993}]{L-galaxies}
{Kauffmann} G.,  {White} S.~D.~M.,   {Guiderdoni} B.,  1993, \mn@doi [\mnras] {10.1093/mnras/264.1.201}, \href {https://ui.adsabs.harvard.edu/abs/1993MNRAS.264..201K} {264, 201}

\bibitem[\protect\citeauthoryear{{Kaviraj} et~al.,}{{Kaviraj} et~al.}{2017}]{Horizon-AGN}
{Kaviraj} S.,  et~al., 2017, \mn@doi [\mnras] {10.1093/mnras/stx126}, \href {https://ui.adsabs.harvard.edu/abs/2017MNRAS.467.4739K} {467, 4739}

\bibitem[\protect\citeauthoryear{{Kim} \& {Ostriker}}{{Kim} \& {Ostriker}}{2017}]{Kim_15}
{Kim} C.-G.,  {Ostriker} E.~C.,  2017, \mn@doi [\apj] {10.3847/1538-4357/aa8599}, \href {https://ui.adsabs.harvard.edu/abs/2017ApJ...846..133K} {846, 133}

\bibitem[\protect\citeauthoryear{{Klessen} \& {Glover}}{{Klessen} \& {Glover}}{2023}]{PopIII_review}
{Klessen} R.~S.,  {Glover} S. C.~O.,  2023, \mn@doi [\araa] {10.1146/annurev-astro-071221-053453}, \href {https://ui.adsabs.harvard.edu/abs/2023ARA&A..61...65K} {61, 65}

\bibitem[\protect\citeauthoryear{{Kravtsov} \& {Manwadkar}}{{Kravtsov} \& {Manwadkar}}{2022}]{Grumpy_model}
{Kravtsov} A.,  {Manwadkar} V.,  2022, \mn@doi [\mnras] {10.1093/mnras/stac1439}, \href {https://ui.adsabs.harvard.edu/abs/2022MNRAS.514.2667K} {514, 2667}

\bibitem[\protect\citeauthoryear{{Kroupa}}{{Kroupa}}{2001}]{Kroupa_01}
{Kroupa} P.,  2001, \mn@doi [\mnras] {10.1046/j.1365-8711.2001.04022.x}, \href {https://ui.adsabs.harvard.edu/abs/2001MNRAS.322..231K} {322, 231}

\bibitem[\protect\citeauthoryear{{Krumholz}, {McKee}  \& {Bland-Hawthorn}}{{Krumholz} et~al.}{2019}]{Krumholz_19}
{Krumholz} M.~R.,  {McKee} C.~F.,   {Bland-Hawthorn} J.,  2019, \mn@doi [\araa] {10.1146/annurev-astro-091918-104430}, \href {https://ui.adsabs.harvard.edu/abs/2019ARA&A..57..227K} {57, 227}

\bibitem[\protect\citeauthoryear{{Lovell}, {Frenk}, {Eke}, {Jenkins}, {Gao}  \& {Theuns}}{{Lovell} et~al.}{2014}]{Lovell_14}
{Lovell} M.~R.,  {Frenk} C.~S.,  {Eke} V.~R.,  {Jenkins} A.,  {Gao} L.,   {Theuns} T.,  2014, \mn@doi [\mnras] {10.1093/mnras/stt2431}, \href {https://ui.adsabs.harvard.edu/abs/2014MNRAS.439..300L} {439, 300}

\bibitem[\protect\citeauthoryear{{Ludlow}, {Bose}, {Angulo}, {Wang}, {Hellwing}, {Navarro}, {Cole}  \& {Frenk}}{{Ludlow} et~al.}{2016}]{Ludlow_16}
{Ludlow} A.~D.,  {Bose} S.,  {Angulo} R.~E.,  {Wang} L.,  {Hellwing} W.~A.,  {Navarro} J.~F.,  {Cole} S.,   {Frenk} C.~S.,  2016, \mn@doi [\mnras] {10.1093/mnras/stw1046}, \href {https://ui.adsabs.harvard.edu/abs/2016MNRAS.460.1214L} {460, 1214}

\bibitem[\protect\citeauthoryear{{Lupi}, {Haiman}  \& {Volonteri}}{{Lupi} et~al.}{2021}]{Lupi_21}
{Lupi} A.,  {Haiman} Z.,   {Volonteri} M.,  2021, \mn@doi [\mnras] {10.1093/mnras/stab692}, \href {https://ui.adsabs.harvard.edu/abs/2021MNRAS.503.5046L} {503, 5046}

\bibitem[\protect\citeauthoryear{{Machacek}, {Bryan}  \& {Abel}}{{Machacek} et~al.}{2001}]{Machacek_01}
{Machacek} M.~E.,  {Bryan} G.~L.,   {Abel} T.,  2001, \mn@doi [\apj] {10.1086/319014}, \href {https://ui.adsabs.harvard.edu/abs/2001ApJ...548..509M} {548, 509}

\bibitem[\protect\citeauthoryear{{Manwadkar} \& {Kravtsov}}{{Manwadkar} \& {Kravtsov}}{2022}]{Grumpy_consttraint}
{Manwadkar} V.,  {Kravtsov} A.~V.,  2022, \mn@doi [\mnras] {10.1093/mnras/stac2452}, \href {https://ui.adsabs.harvard.edu/abs/2022MNRAS.516.3944M} {516, 3944}

\bibitem[\protect\citeauthoryear{{Marsh}}{{Marsh}}{2015}]{Marsh_15}
{Marsh} D. J.~E.,  2015, \mn@doi [arXiv e-prints] {10.48550/arXiv.1510.07633}, \href {https://ui.adsabs.harvard.edu/abs/2015arXiv151007633M} {p. arXiv:1510.07633}

\bibitem[\protect\citeauthoryear{{May} \& {Springel}}{{May} \& {Springel}}{2023}]{May_23}
{May} S.,  {Springel} V.,  2023, \mn@doi [\mnras] {10.1093/mnras/stad2031}, \href {https://ui.adsabs.harvard.edu/abs/2023MNRAS.524.4256M} {524, 4256}

\bibitem[\protect\citeauthoryear{{Moster}, {Naab}  \& {White}}{{Moster} et~al.}{2013}]{Moster_13}
{Moster} B.~P.,  {Naab} T.,   {White} S. D.~M.,  2013, \mn@doi [\mnras] {10.1093/mnras/sts261}, \href {https://ui.adsabs.harvard.edu/abs/2013MNRAS.428.3121M} {428, 3121}

\bibitem[\protect\citeauthoryear{{Naab} \& {Ostriker}}{{Naab} \& {Ostriker}}{2017}]{Naab_17}
{Naab} T.,  {Ostriker} J.~P.,  2017, \mn@doi [\araa] {10.1146/annurev-astro-081913-040019}, \href {https://ui.adsabs.harvard.edu/abs/2017ARA&A..55...59N} {55, 59}

\bibitem[\protect\citeauthoryear{{Nadler}}{{Nadler}}{2025}]{Nadler_25}
{Nadler} E.~O.,  2025, \mn@doi [\apjl] {10.3847/2041-8213/adbc6e}, \href {https://ui.adsabs.harvard.edu/abs/2025ApJ...983L..23N} {983, L23}

\bibitem[\protect\citeauthoryear{{Nadler} et~al.,}{{Nadler} et~al.}{2020}]{Nadler_20}
{Nadler} E.~O.,  et~al., 2020, \mn@doi [\apj] {10.3847/1538-4357/ab846a}, \href {https://ui.adsabs.harvard.edu/abs/2020ApJ...893...48N} {893, 48}

\bibitem[\protect\citeauthoryear{{Nadler} et~al.,}{{Nadler} et~al.}{2021}]{Nadler_21_cons}
{Nadler} E.~O.,  et~al., 2021, \mn@doi [\prl] {10.1103/PhysRevLett.126.091101}, \href {https://ui.adsabs.harvard.edu/abs/2021PhRvL.126i1101N} {126, 091101}

\bibitem[\protect\citeauthoryear{{Navarro}, {Frenk}  \& {White}}{{Navarro} et~al.}{1997}]{NFW}
{Navarro} J.~F.,  {Frenk} C.~S.,   {White} S. D.~M.,  1997, \mn@doi [\apj] {10.1086/304888}, \href {https://ui.adsabs.harvard.edu/abs/1997ApJ...490..493N} {490, 493}

\bibitem[\protect\citeauthoryear{{Nebrin}, {Giri}  \& {Mellema}}{{Nebrin} et~al.}{2023}]{Nebrin_23}
{Nebrin} O.,  {Giri} S.~K.,   {Mellema} G.,  2023, \mn@doi [\mnras] {10.1093/mnras/stad1852}, \href {https://ui.adsabs.harvard.edu/abs/2023MNRAS.524.2290N} {524, 2290}

\bibitem[\protect\citeauthoryear{{Nelson} et~al.,}{{Nelson} et~al.}{2019}]{TNG1}
{Nelson} D.,  et~al., 2019, \mn@doi [\mnras] {10.1093/mnras/stz2306}, \href {https://ui.adsabs.harvard.edu/abs/2019MNRAS.490.3234N} {490, 3234}

\bibitem[\protect\citeauthoryear{{Newton}, {Cautun}, {Jenkins}, {Frenk}  \& {Helly}}{{Newton} et~al.}{2018}]{Newton_18}
{Newton} O.,  {Cautun} M.,  {Jenkins} A.,  {Frenk} C.~S.,   {Helly} J.~C.,  2018, \mn@doi [\mnras] {10.1093/mnras/sty1085}, \href {https://ui.adsabs.harvard.edu/abs/2018MNRAS.479.2853N} {479, 2853}

\bibitem[\protect\citeauthoryear{{Pace}}{{Pace}}{2024}]{Pace_24}
{Pace} A.~B.,  2024, \mn@doi [arXiv e-prints] {10.48550/arXiv.2411.07424}, \href {https://ui.adsabs.harvard.edu/abs/2024arXiv241107424P} {p. arXiv:2411.07424}

\bibitem[\protect\citeauthoryear{{Pereira-Wilson}, {Navarro}, {Ben{\'\i}tez-Llambay}  \& {Santos-Santos}}{{Pereira-Wilson} et~al.}{2023}]{Apostle_sf_thresh}
{Pereira-Wilson} M.,  {Navarro} J.~F.,  {Ben{\'\i}tez-Llambay} A.,   {Santos-Santos} I.,  2023, \mn@doi [\mnras] {10.1093/mnras/stac3633}, \href {https://ui.adsabs.harvard.edu/abs/2023MNRAS.519.1425P} {519, 1425}

\bibitem[\protect\citeauthoryear{{Pillepich} et~al.,}{{Pillepich} et~al.}{2019}]{TNG2}
{Pillepich} A.,  et~al., 2019, \mn@doi [\mnras] {10.1093/mnras/stz2338}, \href {https://ui.adsabs.harvard.edu/abs/2019MNRAS.490.3196P} {490, 3196}

\bibitem[\protect\citeauthoryear{{Planck Collaboration} et~al.,}{{Planck Collaboration} et~al.}{2014}]{Planck_13}
{Planck Collaboration} et~al., 2014, \mn@doi [\aap] {10.1051/0004-6361/201321591}, \href {https://ui.adsabs.harvard.edu/abs/2014A&A...571A..16P} {571, A16}

\bibitem[\protect\citeauthoryear{{Ploeckinger} \& {Schaye}}{{Ploeckinger} \& {Schaye}}{2020}]{Ploeckinger_20}
{Ploeckinger} S.,  {Schaye} J.,  2020, \mn@doi [\mnras] {10.1093/mnras/staa2172}, \href {https://ui.adsabs.harvard.edu/abs/2020MNRAS.497.4857P} {497, 4857}

\bibitem[\protect\citeauthoryear{{Ploeckinger}, {Richings}, {Schaye}, {Trayford}, {Schaller}  \& {Chaikin}}{{Ploeckinger} et~al.}{2025}]{Ploeckinger_25}
{Ploeckinger} S.,  {Richings} A.~J.,  {Schaye} J.,  {Trayford} J.~W.,  {Schaller} M.,   {Chaikin} E.,  2025, \mn@doi [arXiv e-prints] {10.48550/arXiv.2506.15773}, \href {https://ui.adsabs.harvard.edu/abs/2025arXiv250615773P} {p. arXiv:2506.15773}

\bibitem[\protect\citeauthoryear{{Portinari}, {Chiosi}  \& {Bressan}}{{Portinari} et~al.}{1998}]{Portinari_98}
{Portinari} L.,  {Chiosi} C.,   {Bressan} A.,  1998, \mn@doi [\aap] {10.48550/arXiv.astro-ph/9711337}, \href {https://ui.adsabs.harvard.edu/abs/1998A&A...334..505P} {334, 505}

\bibitem[\protect\citeauthoryear{{Rahmati}, {Pawlik}, {Rai{\v{c}}evi{\'c}}  \& {Schaye}}{{Rahmati} et~al.}{2013}]{Rahmati_13}
{Rahmati} A.,  {Pawlik} A.~H.,  {Rai{\v{c}}evi{\'c}} M.,   {Schaye} J.,  2013, \mn@doi [\mnras] {10.1093/mnras/stt066}, \href {https://ui.adsabs.harvard.edu/abs/2013MNRAS.430.2427R} {430, 2427}

\bibitem[\protect\citeauthoryear{{Revaz} \& {Jablonka}}{{Revaz} \& {Jablonka}}{2018}]{Gear}
{Revaz} Y.,  {Jablonka} P.,  2018, \mn@doi [\aap] {10.1051/0004-6361/201832669}, \href {https://ui.adsabs.harvard.edu/abs/2018A&A...616A..96R} {616, A96}

\bibitem[\protect\citeauthoryear{{Rey}, {Pontzen}, {Agertz}, {Orkney}, {Read}, {Saintonge}  \& {Pedersen}}{{Rey} et~al.}{2019}]{Edge_1}
{Rey} M.~P.,  {Pontzen} A.,  {Agertz} O.,  {Orkney} M. D.~A.,  {Read} J.~I.,  {Saintonge} A.,   {Pedersen} C.,  2019, \mn@doi [\apjl] {10.3847/2041-8213/ab53dd}, \href {https://ui.adsabs.harvard.edu/abs/2019ApJ...886L...3R} {886, L3}

\bibitem[\protect\citeauthoryear{{Rey}, {Pontzen}, {Agertz}, {Orkney}, {Read}  \& {Rosdahl}}{{Rey} et~al.}{2020}]{Rey_20}
{Rey} M.~P.,  {Pontzen} A.,  {Agertz} O.,  {Orkney} M. D.~A.,  {Read} J.~I.,   {Rosdahl} J.,  2020, \mn@doi [\mnras] {10.1093/mnras/staa1640}, \href {https://ui.adsabs.harvard.edu/abs/2020MNRAS.497.1508R} {497, 1508}

\bibitem[\protect\citeauthoryear{{Rey} et~al.,}{{Rey} et~al.}{2025}]{Edge_2}
{Rey} M.~P.,  et~al., 2025, \mn@doi [\mnras] {10.1093/mnras/staf1058}, \href {https://ui.adsabs.harvard.edu/abs/2025MNRAS.541.1195R} {541, 1195}

\bibitem[\protect\citeauthoryear{{Richings}, {Schaye}  \& {Oppenheimer}}{{Richings} et~al.}{2014a}]{CHIMES_1}
{Richings} A.~J.,  {Schaye} J.,   {Oppenheimer} B.~D.,  2014a, \mn@doi [\mnras] {10.1093/mnras/stu525}, \href {https://ui.adsabs.harvard.edu/abs/2014MNRAS.440.3349R} {440, 3349}

\bibitem[\protect\citeauthoryear{{Richings}, {Schaye}  \& {Oppenheimer}}{{Richings} et~al.}{2014b}]{CHIMES_2}
{Richings} A.~J.,  {Schaye} J.,   {Oppenheimer} B.~D.,  2014b, \mn@doi [\mnras] {10.1093/mnras/stu525}, \href {https://ui.adsabs.harvard.edu/abs/2014MNRAS.440.3349R} {440, 3349}

\bibitem[\protect\citeauthoryear{{Sawala} et~al.,}{{Sawala} et~al.}{2016}]{APOSTLE_1}
{Sawala} T.,  et~al., 2016, \mn@doi [\mnras] {10.1093/mnras/stw145}, \href {https://ui.adsabs.harvard.edu/abs/2016MNRAS.457.1931S} {457, 1931}

\bibitem[\protect\citeauthoryear{{Schauer}, {Glover}, {Klessen}  \& {Ceverino}}{{Schauer} et~al.}{2019}]{Schauer_19}
{Schauer} A. T.~P.,  {Glover} S. C.~O.,  {Klessen} R.~S.,   {Ceverino} D.,  2019, \mn@doi [\mnras] {10.1093/mnras/stz013}, \href {https://ui.adsabs.harvard.edu/abs/2019MNRAS.484.3510S} {484, 3510}

\bibitem[\protect\citeauthoryear{{Schauer}, {Glover}, {Klessen}  \& {Clark}}{{Schauer} et~al.}{2021}]{Schauer_21}
{Schauer} A. T.~P.,  {Glover} S. C.~O.,  {Klessen} R.~S.,   {Clark} P.,  2021, \mn@doi [\mnras] {10.1093/mnras/stab1953}, \href {https://ui.adsabs.harvard.edu/abs/2021MNRAS.507.1775S} {507, 1775}

\bibitem[\protect\citeauthoryear{{Schauer}, {Boylan-Kolchin}, {Colston}, {Sameie}, {Bromm}, {Bullock}  \& {Wetzel}}{{Schauer} et~al.}{2023}]{Schauer_23}
{Schauer} A. T.~P.,  {Boylan-Kolchin} M.,  {Colston} K.,  {Sameie} O.,  {Bromm} V.,  {Bullock} J.~S.,   {Wetzel} A.,  2023, \mn@doi [\apj] {10.3847/1538-4357/accc2c}, \href {https://ui.adsabs.harvard.edu/abs/2023ApJ...950...20S} {950, 20}

\bibitem[\protect\citeauthoryear{{Schaye} et~al.,}{{Schaye} et~al.}{2015}]{EAGLE_1}
{Schaye} J.,  et~al., 2015, \mn@doi [\mnras] {10.1093/mnras/stu2058}, \href {https://ui.adsabs.harvard.edu/abs/2015MNRAS.446..521S} {446, 521}

\bibitem[\protect\citeauthoryear{{Schaye} et~al.,}{{Schaye} et~al.}{2025}]{Colibre}
{Schaye} J.,  et~al., 2025, \mn@doi [arXiv e-prints] {10.48550/arXiv.2508.21126}, \href {https://ui.adsabs.harvard.edu/abs/2025arXiv250821126S} {p. arXiv:2508.21126}

\bibitem[\protect\citeauthoryear{{Simon}}{{Simon}}{2019}]{dwarf_review}
{Simon} J.~D.,  2019, \mn@doi [\araa] {10.1146/annurev-astro-091918-104453}, \href {https://ui.adsabs.harvard.edu/abs/2019ARA&A..57..375S} {57, 375}

\bibitem[\protect\citeauthoryear{{Skinner} \& {Wise}}{{Skinner} \& {Wise}}{2020}]{Skinner_20}
{Skinner} D.,  {Wise} J.~H.,  2020, \mn@doi [\mnras] {10.1093/mnras/staa139}, \href {https://ui.adsabs.harvard.edu/abs/2020MNRAS.492.4386S} {492, 4386}

\bibitem[\protect\citeauthoryear{{Solomon}}{{Solomon}}{1965}]{Solomon_65}
{Solomon} P.~M.,  1965, The University of Wisconsin-Madison

\bibitem[\protect\citeauthoryear{{Springel}}{{Springel}}{2010}]{AREOP_1}
{Springel} V.,  2010, \mn@doi [\mnras] {10.1111/j.1365-2966.2009.15715.x}, \href {https://ui.adsabs.harvard.edu/abs/2010MNRAS.401..791S} {401, 791}

\bibitem[\protect\citeauthoryear{{Springel}, {White}, {Tormen}  \& {Kauffmann}}{{Springel} et~al.}{2001}]{Subfind}
{Springel} V.,  {White} S. D.~M.,  {Tormen} G.,   {Kauffmann} G.,  2001, \mn@doi [\mnras] {10.1046/j.1365-8711.2001.04912.x}, \href {https://ui.adsabs.harvard.edu/abs/2001MNRAS.328..726S} {328, 726}

\bibitem[\protect\citeauthoryear{{Springel} et~al.,}{{Springel} et~al.}{2008}]{Aquarious}
{Springel} V.,  et~al., 2008, \mn@doi [\mnras] {10.1111/j.1365-2966.2008.14066.x}, \href {https://ui.adsabs.harvard.edu/abs/2008MNRAS.391.1685S} {391, 1685}

\bibitem[\protect\citeauthoryear{{Srisawat} et~al.,}{{Srisawat} et~al.}{2013}]{Srisawat_13}
{Srisawat} C.,  et~al., 2013, \mn@doi [\mnras] {10.1093/mnras/stt1545}, \href {https://ui.adsabs.harvard.edu/abs/2013MNRAS.436..150S} {436, 150}

\bibitem[\protect\citeauthoryear{{Stafford}, {Brown}, {McCarthy}, {Font}, {Robertson}  \& {Poole-McKenzie}}{{Stafford} et~al.}{2020}]{Stafford_20}
{Stafford} S.~G.,  {Brown} S.~T.,  {McCarthy} I.~G.,  {Font} A.~S.,  {Robertson} A.,   {Poole-McKenzie} R.,  2020, \mn@doi [\mnras] {10.1093/mnras/staa2059}, \href {https://ui.adsabs.harvard.edu/abs/2020MNRAS.497.3809S} {497, 3809}

\bibitem[\protect\citeauthoryear{{Stecher} \& {Williams}}{{Stecher} \& {Williams}}{1967}]{Stecher_67}
{Stecher} T.~P.,  {Williams} D.~A.,  1967, \mn@doi [\apjl] {10.1086/180047}, \href {https://ui.adsabs.harvard.edu/abs/1967ApJ...149L..29S} {149, L29}

\bibitem[\protect\citeauthoryear{{Sukhbold}, {Ertl}, {Woosley}, {Brown}  \& {Janka}}{{Sukhbold} et~al.}{2016}]{Sukhbold_16}
{Sukhbold} T.,  {Ertl} T.,  {Woosley} S.~E.,  {Brown} J.~M.,   {Janka} H.~T.,  2016, \mn@doi [\apj] {10.3847/0004-637X/821/1/38}, \href {https://ui.adsabs.harvard.edu/abs/2016ApJ...821...38S} {821, 38}

\bibitem[\protect\citeauthoryear{{Sureda}, {Brown}, {Fattahi}, {Gutcke}, {Bose}, {Doppel}  \& {Pakmor}}{{Sureda} et~al.}{2025}]{Sureda}
{Sureda} J.,  {Brown} S.~T.,  {Fattahi} A.,  {Gutcke} T.,  {Bose} S.,  {Doppel} J.~E.,   {Pakmor} R.,  2025, \mn@doi [arXiv e-prints] {10.48550/arXiv.2511.10582}, \href {https://ui.adsabs.harvard.edu/abs/2025arXiv251110582S} {p. arXiv:2511.10582}

\bibitem[\protect\citeauthoryear{{Taylor} et~al.,}{{Taylor} et~al.}{2025}]{Taylor_25}
{Taylor} E.~D.,  et~al., 2025, \mn@doi [arXiv e-prints] {10.48550/arXiv.2509.09582}, \href {https://ui.adsabs.harvard.edu/abs/2025arXiv250909582T} {p. arXiv:2509.09582}

\bibitem[\protect\citeauthoryear{{Tolstoy}, {Hill}  \& {Tosi}}{{Tolstoy} et~al.}{2009}]{Tolstoy_09}
{Tolstoy} E.,  {Hill} V.,   {Tosi} M.,  2009, \mn@doi [\araa] {10.1146/annurev-astro-082708-101650}, \href {https://ui.adsabs.harvard.edu/abs/2009ARA&A..47..371T} {47, 371}

\bibitem[\protect\citeauthoryear{{Tremmel}, {Karcher}, {Governato}, {Volonteri}, {Quinn}, {Pontzen}, {Anderson}  \& {Bellovary}}{{Tremmel} et~al.}{2017}]{Romulus}
{Tremmel} M.,  {Karcher} M.,  {Governato} F.,  {Volonteri} M.,  {Quinn} T.~R.,  {Pontzen} A.,  {Anderson} L.,   {Bellovary} J.,  2017, \mn@doi [\mnras] {10.1093/mnras/stx1160}, \href {https://ui.adsabs.harvard.edu/abs/2017MNRAS.470.1121T} {470, 1121}

\bibitem[\protect\citeauthoryear{{Trenti} \& {Stiavelli}}{{Trenti} \& {Stiavelli}}{2009}]{Trenti_09}
{Trenti} M.,  {Stiavelli} M.,  2009, \mn@doi [\apj] {10.1088/0004-637X/694/2/879}, \href {https://ui.adsabs.harvard.edu/abs/2009ApJ...694..879T} {694, 879}

\bibitem[\protect\citeauthoryear{{Tseliakhovich} \& {Hirata}}{{Tseliakhovich} \& {Hirata}}{2010}]{Tseliakhovich}
{Tseliakhovich} D.,  {Hirata} C.,  2010, \mn@doi [\prd] {10.1103/PhysRevD.82.083520}, \href {https://ui.adsabs.harvard.edu/abs/2010PhRvD..82h3520T} {82, 083520}

\bibitem[\protect\citeauthoryear{{Vogelsberger}, {Marinacci}, {Torrey}  \& {Puchwein}}{{Vogelsberger} et~al.}{2020}]{Vogelsberger_review}
{Vogelsberger} M.,  {Marinacci} F.,  {Torrey} P.,   {Puchwein} E.,  2020, \mn@doi [Nature Reviews Physics] {10.1038/s42254-019-0127-2}, \href {https://ui.adsabs.harvard.edu/abs/2020NatRP...2...42V} {2, 42}

\bibitem[\protect\citeauthoryear{{Wang}, {Cooper}, {Bose}, {Frenk}  \& {Hellwing}}{{Wang} et~al.}{2023}]{Ghost_galaxies}
{Wang} C.-W.,  {Cooper} A.~P.,  {Bose} S.,  {Frenk} C.~S.,   {Hellwing} W.~A.,  2023, \mn@doi [\apj] {10.3847/1538-4357/ad011d}, \href {https://ui.adsabs.harvard.edu/abs/2023ApJ...958..166W} {958, 166}

\bibitem[\protect\citeauthoryear{{Weinberger}, {Springel}  \& {Pakmor}}{{Weinberger} et~al.}{2020}]{AREPO_2}
{Weinberger} R.,  {Springel} V.,   {Pakmor} R.,  2020, \mn@doi [\apjs] {10.3847/1538-4365/ab908c}, \href {https://ui.adsabs.harvard.edu/abs/2020ApJS..248...32W} {248, 32}

\bibitem[\protect\citeauthoryear{{Weisz}, {Dolphin}, {Skillman}, {Holtzman}, {Gilbert}, {Dalcanton}  \& {Williams}}{{Weisz} et~al.}{2014}]{Weisz_14}
{Weisz} D.~R.,  {Dolphin} A.~E.,  {Skillman} E.~D.,  {Holtzman} J.,  {Gilbert} K.~M.,  {Dalcanton} J.~J.,   {Williams} B.~F.,  2014, \mn@doi [\apj] {10.1088/0004-637X/789/2/147}, \href {https://ui.adsabs.harvard.edu/abs/2014ApJ...789..147W} {789, 147}

\bibitem[\protect\citeauthoryear{{Wheeler} et~al.,}{{Wheeler} et~al.}{2019}]{Wheeler}
{Wheeler} C.,  et~al., 2019, \mn@doi [\mnras] {10.1093/mnras/stz2887}, \href {https://ui.adsabs.harvard.edu/abs/2019MNRAS.490.4447W} {490, 4447}

\bibitem[\protect\citeauthoryear{{Wise}, {Abel}, {Turk}, {Norman}  \& {Smith}}{{Wise} et~al.}{2012}]{Wise_12}
{Wise} J.~H.,  {Abel} T.,  {Turk} M.~J.,  {Norman} M.~L.,   {Smith} B.~D.,  2012, \mn@doi [\mnras] {10.1111/j.1365-2966.2012.21809.x}, \href {https://ui.adsabs.harvard.edu/abs/2012MNRAS.427..311W} {427, 311}

\bibitem[\protect\citeauthoryear{{Zhou}, {Zhu}, {Yang}, {Yu}, {Yuan}, {Jiang}  \& {Xi}}{{Zhou} et~al.}{2023}]{Zhou_23}
{Zhou} R.,  {Zhu} M.,  {Yang} Y.,  {Yu} H.,  {Yuan} L.,  {Jiang} P.,   {Xi} W.,  2023, \mn@doi [\apj] {10.3847/1538-4357/acdcf5}, \href {https://ui.adsabs.harvard.edu/abs/2023ApJ...952..130Z} {952, 130}

\makeatother
\end{thebibliography}

%\begin{thebibliography}{99}
%\end{thebibliography}

%%%%%%%%%%%%%%%%%%%%%%%%%%%%%%%%%%%%%%%%%%%%%%%%%%

%%%%%%%%%%%%%%%%% APPENDICES %%%%%%%%%%%%%%%%%%%%%

\appendix

\section{Sample selection} \label{sec:suite_stats}

\begin{figure}
    \centering
    \includegraphics[width=1.0\linewidth]{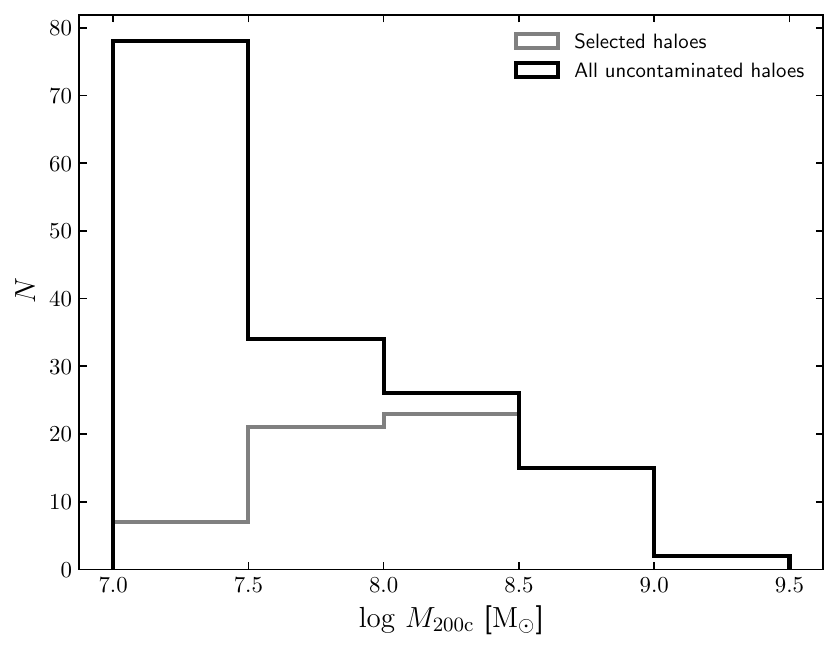}
    \caption{Distribution of the present day halo masses within the new simulation suite. The haloes centred on for resimulation are shown in gray. All uncontaminated haloes with \Mhalo$>10^7$\Msun are shown in black.}
    \label{fig:Halo_suite}
\end{figure}

\begin{table*}
\centering
\caption{The halo and galaxy properties for the new simulation suite. The simulation name is shown in column (b), column (a) shows the reference name used in the paper; the majority of haloes are not individually referenced. Column (c) and (d) present a measure of size of the main halo today, specifically the mass, \Mhalo, and the maximum circular velocity, respectively. For simplicity these values are taken from the stronger LWB, however \Mhalo and $V_{\mathrm{max}}$ change by $\lesssim 10 \%$ between the simulation. In columns (e) to (h) we present properties of the stellar component of the haloes in the two different LWB, showing the total stellar mass today and the redshift when the halo first forms stars.}
\label{tab:suite}
\begin{tabular}{llllllll}
\hline
\multicolumn{4}{l}{ } & \multicolumn{2}{l}{Stronger LWB} & \multicolumn{2}{c}{Weaker LWB} \\
(a) Reference & (b) Halo  Name & (c) $M_{\mathrm{200c}}$ {[}$\mathrm{M_{\odot}}${]} & (d) $V_{\mathrm{max}}$ [$\mathrm{km/s}$] & (e) $M_{\ast}$ {[}$\mathrm{M_{\odot}}${]} & (f) $z_{\mathrm{form}}$ & (g) $M_{\ast}$ {[}$\mathrm{M_{\odot}}${]} & (h)$z_{\mathrm{form}}$ \\ \hline
              	& halo 000  & 7.58$\times 10^8$                          	& 17.1           	& 3.72$\times 10^4$                        	& 25.9                   	& 3.38$\times 10^5$                            	& 26.0                     	\\
              	& halo 001  & 7.19$\times 10^8$                          	& 16.2           	& 1.48$\times 10^5$                        	& 25.7  & 1.50$\times 10^5$                            	& 25.8                     	\\
              	& halo 005  & 6.43$\times 10^8$                          	& 16.9           	& 2.66$\times 10^4$                        	& 22.1                   	& 2.32$\times 10^5$                            	& 22.3                     	\\
              	& halo 023  & 3.76$\times 10^8$                          	& 13.1           	& 3.98$\times 10^3$                        	& 23.0                   	& 7.58$\times 10^4$                            	& 23.2                     	\\
              	& halo 009  & 4.51$\times 10^8$                          	& 14.2           	& 3.61$\times 10^3$                        	& 20.7                   	& 5.71$\times 10^4$                            	& 20.8                     	\\
$\mathrm{B}$  	& halo 007  & 6.33$\times 10^8$                          	& 16.3           	& 1.14$\times 10^5$                        	& 24.4                   	& 2.17$\times 10^5$                            	& 24.4                     	\\
              	& halo 011  & 4.51$\times 10^8$                          	& 13.1           	& 1.34$\times 10^3$                        	& 20.0                  	& 5.03$\times 10^4$                            	& 21.1                     	\\
              	& halo 034  & 1.67$\times 10^8$                          	& 9.52           	& 0                                        	& --                     	& 5.20$\times 10^4$                            	& 18.0                     	\\
              	& halo 033  & 1.81$\times 10^8$                          	& 10.7           	& 0                                        	& --                     	& 5.76$\times 10^4$                            	& 18.4                     	\\
              	& halo 028  & 2.11$\times 10^8$                          	& 9.82           	& 0                                        	& --                     	& 2.49$\times 10^4$                            	& 18.1                     	\\
              	& halo 032  & 2.02$\times 10^8$                          	& 10.6           	& 0                                        	& --                     	& 3.54$\times 10^4$                            	& 16.0                     	\\
              	& halo 029  & 1.94$\times 10^8$                          	& 10.4           	& 0                                        	& --                     	& 3.61$\times 10^4$                            	& 16.9                     	\\
              	& halo 030  & 2.15$\times 10^8$                          	& 9.92           	& 0                                        	& --                     	& 1.13$\times 10^4$                            	& 18.8                     	\\
              	& halo 045  & 1.22$\times 10^8$                          	& 8.92           	& 0                                        	& --                     	& 3.55$\times 10^4$                            	& 17.5                     	\\
$\mathrm{C}$  	& halo 018  & 1.74$\times 10^8$                          	& 9.63           	& 4.22$\times 10^3$                        	& 20.4                   	& 3.30$\times 10^4$                            	& 20.6                     	\\
              	& halo 036  & 1.75$\times 10^8$                          	& 10.6           	& 0                                        	& --                     	& 4.44$\times 10^4$                            	& 19.2                     	\\
              	& halo 058  & 9.01$\times 10^7$                          	& 8.78           	& 0                                        	& --                     	& 3.56$\times 10^4$                            	& 18.0                     	\\
              	& halo 078  & 5.46$\times 10^7$                          	& 6.58           	& 0                                        	& --                     	& 3.97$\times 10^2$                            	& 9.23                     	\\
              	& halo 100  & 3.70$\times 10^7$                          	& 5.96           	& 0                                        	& --                     	& 7.34$\times 10^3$                            	& 13.9                     	\\
              	& halo 060  & 7.83$\times 10^7$                          	& 7.79           	& 0                                        	& --                     	& 3.57$\times 10^4$                            	& 12.4                     	\\
              	& halo 074  & 6.10$\times 10^7$                          	& 7.10           	& 0                                        	& --                     	& 1.84$\times 10^4$                            	& 17.5                     	\\
              	& halo 057  & 7.51$\times 10^7$                          	& 8.03           	& 3.21$\times 10^3$                        	& 20.7                   	& 2.05$\times 10^4$                            	& 20.9                     	\\
              	& halo 101  & 3.64$\times 10^7$                          	& 5.86           	& 0                                        	& --                     	& 7.18$\times 10^3$                            	& 8.49                     	\\
              	& halo 056  & 8.60$\times 10^7$                          	& 7.91           	& 0                                        	& --                     	& 1.63$\times 10^4$                            	& 18.0                     	\\
              	& halo 099  & 3.83$\times 10^7$                          	& 5.67           	& 0                                        	& --                     	& 6.51$\times 10^3$                            	& 9.98                     	\\
              	& halo 096  & 3.99$\times 10^7$                          	& 5.87           	& 0                                        	& --                     	& 9.42$\times 10^3$                            	& 14.0                     	\\
              	& halo 103  & 3.38$\times 10^7$                          	& 5.96           	& 0                                        	& --                     	& 2.60$\times 10^3$                            	& 13.2                     	\\
              	& halo 114  & 1.74$\times 10^7$                          	& 4.76           	& 0                                        	& --                     	& 0                                            	& --                       	\\
              	& halo 115  & 1.74$\times 10^7$                          	& 4.76           	& 0                                        	& --                     	& 0                                            	& --                       	\\
              	& halo 123  & 2.29$\times 10^7$                          	& 4.78           	& 0                                        	& --                     	& 1.60$\times 10^3$                            	& 11.1                     	\\
              	& halo 127  & 2.21$\times 10^7$                          	& 5.02           	& 0                                        	& --                     	& 1.46$\times 10^3$                            	& 12.8                     	\\
$\mathrm{D}$  	& halo 134  & 1.61$\times 10^7$                          	& 4.81           	& 0                                        	& --                     	& 6.26$\times 10^3$                            	& 9.77                     	\\
              	& halo 135  & 1.61$\times 10^7$                          	& 4.81           	& 0                                        	& --                     	& 6.26$\times 10^3$                            	& 9.77                     	\\
              	& halo 148  & 1.26$\times 10^7$                          	& 4.46           	& 0                                        	& --                     	& 2.72$\times 10^3$                            	& 9.52                     	\\
              	& halo 054  & 9.40$\times 10^7$                          	& 9.25           	& 7.41$\times 10^2$                        	& 19.6                   	& 5.32$\times 10^4$                            	& 20.4                     	\\
              	& halo 050  & 9.79$\times 10^7$                          	& 8.26           	& 0                                        	& --                     	& 1.75$\times 10^4$                            	& 18.2                     	\\
              	& halo 188  & 8.37$\times 10^7$                          	& 8.13           	& 0                                        	& --                     	& 9.45$\times 10^3$                            	& 15.0                     	\\
              	& halo 165  & 9.52$\times 10^7$                          	& 8.40           	& 0                                        	& --                     	& 2.25$\times 10^4$                            	& 15.9                     	\\
              	& halo 199  & 1.09$\times 10^8$                          	& 8.27           	& 0                                        	& --                     	& 1.94$\times 10^4$                            	& 12.7                     	\\
              	& halo 164  & 1.11$\times 10^8$                          	& 9.00           	& 0                                        	& --                     	& 2.54$\times 10^4$                            	& 16.7                     	\\
              	& halo 180  & 1.02$\times 10^8$                          	& 7.87           	& 0                                        	& --                     	& 8.22$\times 10^3$                            	& 12.1                     	\\
              	& halo 163  & 1.19$\times 10^8$                          	& 8.32           	& 0                                        	& --                     	& 1.19$\times 10^4$                            	& 14.5                     	\\
              	& halo 173  & 1.24$\times 10^8$                          	& 8.92           	& 0                                        	& --                     	& 3.62$\times 10^4$                            	& 15.6                     	\\
              	& halo 185  & 1.32$\times 10^8$                          	& 9.04           	& 0                                        	& --                     	& 1.66$\times 10^4$                            	& 12.6                     	\\
              	& halo 175  & 1.88$\times 10^8$                          	& 9.70           	& 0                                        	& --                     	& 3.20$\times 10^4$                            	& 18.8                     	\\
              	& halo 176  & 2.97$\times 10^8$                          	& 11.6           	& 0                                        	& --                     	& 4.46$\times 10^4$                            	& 16.8                     	\\
              	& halo 187  & 2.33$\times 10^8$                          	& 10.0           	& 0                                        	& --                     	& 2.43$\times 10^4$                            	& 19.3                     	\\
              	& halo 161  & 4.35$\times 10^8$                          	& 14.0           	& 7.49$\times 10^3$                        	& 21.0                   	& 1.09$\times 10^5$                            	& 21.1                     	\\
              	& halo 168  & 4.22$\times 10^8$                          	& 9.82           	& 0                                        	& --                     	& 2.54$\times 10^4$                            	& 20.6                     	\\
              	& halo 184  & 4.97$\times 10^8$                          	& 15.4           	& 6.06$\times 10^3$                        	& 23.1                   	& 1.36$\times 10^5$                            	& 23.2                     	\\
              	& halo 171  & 5.50$\times 10^8$                          	& 15.6           	& 2.22$\times 10^4$                        	& 23.7                   	& 1.76$\times 10^5$                            	& 23.9                     	\\
              	& halo 190  & 7.38$\times 10^7$                          	& 5.69           	& 0                                        	& --                     	& 9.98$\times 10^3$                            	& 21.4                     	\\
              	& halo 201  & 8.32$\times 10^7$                          	& 6.65           	& 0                                        	& --                     	& 1.30$\times 10^4$                            	& 16.4                     	\\
              	& halo 203  & 1.19$\times 10^8$                          	& 8.79           	& 0                                        	& --                     	& 1.29$\times 10^4$                            	& 17.0                     	\\
              	& halo 215  & 8.72$\times 10^7$                          	& 8.29           	& 0                                        	& --                     	& 2.39$\times 10^4$                            	& 17.5                     	\\
              	& halo 219  & 1.98$\times 10^8$                          	& 11.0           	& 0                                        	& --                     	& 5.06$\times 10^4$                            	& 17.8                     	\\
              	& halo 220  & 7.17$\times 10^7$                          	& 6.86           	& 0                                        	& --                     	& 5.36$\times 10^3$                            	& 13.6                     	\\
              	\hline
\end{tabular}
\end{table*}

\begin{table*}
\centering
\caption{Continuation of Table.~\ref{tab:suite}. The final three haloes are from the original LYRA sample presented in \protect\cite{LYRA_2}, reran with the new cooling tables.}
\label{tab:suite2}
\begin{tabular}{llllllll}
\hline
\multicolumn{4}{l}{ } & \multicolumn{2}{l}{Stronger LWB} & \multicolumn{2}{l}{Weaker LWB} \\
(a) Reference & (b) Halo  Name & (c) $M_{\mathrm{200c}}$ {[}$\mathrm{M_{\odot}}${]} & (d) $V_{\mathrm{max}}$ [$\mathrm{km/s}$] & (e) $M_{\ast}$ {[}$\mathrm{M_{\odot}}${]} & (f) $z_{\mathrm{form}}$ & (g) $M_{\ast}$ {[}$\mathrm{M_{\odot}}${]} & (h) $z_{\mathrm{form}}$ \\ \hline
              	& halo 222  & 1.90$\times 10^8$                          	& 10.6           	& 1.32$\times 10^3$                        	& 20.0                   	& 3.76$\times 10^4$                            	& 20.4 \\
              	& halo 225  & 1.14$\times 10^8$                          	& 8.65           	& 1.27$\times 10^3$                        	& 19.4                   	& 5.40$\times 10^4$                            	& 19.8                     	\\
              	& halo 227  & 9.75$\times 10^7$                          	& 8.41           	& 0                                        	& --                     	& 3.12$\times 10^4$                            	& 13.6                     	\\
              	& halo 238  & 1.12$\times 10^8$                          	& 9.11           	& 0                                        	& --                     	& 4.16$\times 10^4$                            	& 16.7                     	\\
              	& halo 242  & 8.37$\times 10^7$                          	& 7.60           	& 0                                        	& --                     	& 5.79$\times 10^3$                            	& 12.8                     	\\ \hline
              	& Halo E (Gutcke et al. 21)     	& 7.55$\times 10^8$                          	& 14.9           	& 1.10$\times 10^3$                         	& 20.3                   	& 1.07$\times 10^5$                            	& 20.8                     	\\
              	& Halo B (Gutcke et al. 21)    	& 2.86$\times 10^9$                          	& 23.6           	& 1.10$\times 10^6$                        	& 26.4                   	& 6.30$\times 10^5$                            	& 26.5  \\
             	A & Halo D (Gutcke et al. 21)    	& 1.74$\times 10^9$                          	& 22.6           	& 2.4$\times 10^6$                        	& 25.9                   	& 3.42$\times 10^5$                            	& 24.6
             	\\ \cline{1-8}
\end{tabular}
\end{table*}

In Fig.~\ref{fig:Halo_suite} we present the distribution of present day halo masses for the new simulation suite. Here we show the distribution of masses for haloes that were centred on when generating the zoom-in initial condition (gray), along with all uncontaminated haloes within the suite (black). In total we have run $65$ zoom-in simulations. Including the all uncontaminated systems provides a sample of $155$ haloes with \Mhalo$>10^7$\Msun (the approximate mass where haloes transition from dark to luminous in the weaker LWB).

In Table~\ref{tab:suite} \& ~\ref{tab:suite2} we present some key properties of the main haloes and galaxies within the new simulation suite. This includes their present day halo masses, \Mhalo, maximum circular velocity, along with the stellar mass today and the redshift at which they formed $90\%$ of their stellar mass in the two different LWBs.

\section{Gas phases for halo C} \label{section:appendic_gas}

\begin{figure*}
    \centering
   	 \includegraphics[width=1.0\linewidth]{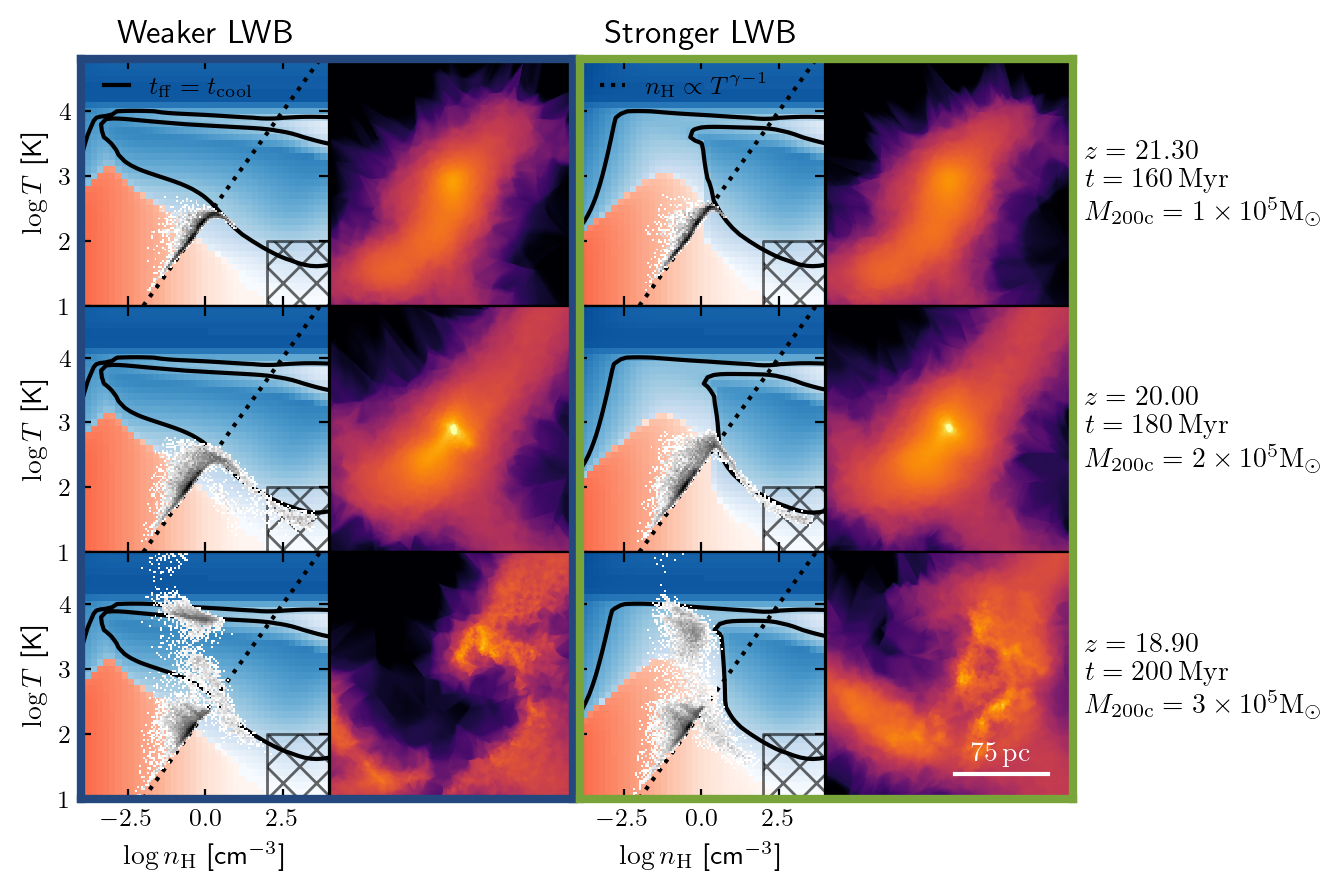}
    \caption{Same as Fig.~\ref{fig:first_stars_halo_134} for Halo B undergoing its first episode of star formation at $z\approx 20$. This occurs at the same time and the behaviour is nearly identical in both LWBs.}
    \label{fig:Gas_first_stars_halo018}
\end{figure*}

Here we study the gas phases of Halo C during its star formation episodes. Halo C is an interesting case where in the weaker LWB it is able to undergo two distinct episodes of star formation before reionisation, along with accreted a number of smaller stellar systems, while in the stronger LWB it is only able to form one burst of stars.

In Fig.~\ref{fig:Gas_first_stars_halo018}, we show the gas phase space and spatial distribution during the first episode of star formation. The figure is constructed the same way as Fig.~\ref{fig:first_stars_halo_134} in the main text.

In the weaker background just before star formation occurs (top left panels), we see that the majority of gas, particularly at lower densities $n_{\mathrm{H}} \lesssim 1 \, \mathrm{cm}^{-3}$, lies on the adiabat and has not had time for significant heating or cooling to occur, and has instead been compressed with little heat transfer. However, at higher densities, we see that the gas at the centre of the halo is beginning to fall off the adiabatic relation and is now cooling efficiently. This occurs roughly when the gas crosses the $t_{\mathrm{ff}} = t_{\mathrm{cool}}$ contour, and can then undergo runaway cooling. In the next snapshot (middle panel), the gas in the centre of the halo now collapses, reaches high enough densities, and cools to low enough temperatures for star formation to occur. There is a combination of diffuse cool gas in the outskirts of the halo, still on the adiabatic relation, with dense cold gas in the centre that is actively star forming. The final snapshot (bottom panel) is chosen just after the first few SN have occurred. Here, the central gas has been shocked and heated due to these SN, which shuts down SF and is in the process of driving a strong outflow that will soon evacuate almost all gas from the halo. The majority of the outflowing gas is at $n_{\mathrm{H}} \approx 1 \, \mathrm{cm^{-3}}$, $T = 10^4\, \mathrm{K}$ at this time.

Focussing now on the stronger LWB (right two columns) we see the same behaviour. Gas initially follows the adiabat, efficiently cooling at $\log \rho/\rm{cm}^{-3} \lesssim 0$, $\log T/\rm{K} \approx 2.5$, then forming stars at the same time and quenching once the first SN occur. We do see minor differences to the shape of distribution of gas in phase space, but this has no meaningful impact on the star formation. At this redshift ($z\sim 20$), the LWBs are different in strength (see Fig.~\ref{fig:LW_background}), leading to moderate differences between the two cooling functions. However, the location the $t_{\mathrm{ff}} = t_{\mathrm{cool}}$ contour (solid black line) crosses the adiabat is at comparable densities and temperatures, and hence there is near identical behaviour between the two models.

%%%%%%%%%%%%%%%%%%%%%%%%%%%%%%%%%%%%%%%%%%%%%%%%%%

% Don't change these lines
\bsp	% typesetting comment
\label{lastpage}
\end{document}